\documentclass[preprint,showpacs,preprintnumbers,amsmath,amssymb]{revtex4-1}
\usepackage{graphicx}
\usepackage{bm}
\usepackage{color}
\usepackage{amssymb}
\usepackage{epsfig}

\begin{document}

\title{Disorder perturbed Flat Bands II: a search for criticality}

\author{Pragya Shukla}
\affiliation{Department of Physics,
 Indian Institute of Technology, Kharagpur, India.}
 
\date{\today}
 \widetext

 \begin{abstract}
 We present a common mathematical formulation of  the level statistics of a disordered tight-binding lattice, with  one or many flat bands in clean limit, in which system specific details enters through a single parameter. The formulation, applicable to both single as well as many particle flat bands, indicates the possibility of two different types of critical statistics: one in weak disorder regime (below a system specific disorder strength) and  insensitive of the disorder-strength, another  in strong disorder regime and occurs at specific critical disorder strengths. The single parametric dependence however relates  the statistics in the two regimes  (notwithstanding  different scattering conditions therein). This also helps in revealing an underlying  universality of the statistics in weakly disordered flat bands, shared by a wide-range of other complex systems irrespective of the origin of their complexity.

\end{abstract}

 

 \maketitle

\section{Introduction}

A dispersion-less band, also referred as a flat band, appears in crystal lattices under subtle  interplay of the system conditions. 
The onset of disorder, say $w$ may lead to violation of these conditions, lifting the degeneracy of the energy levels and changing the nature of the eigenfunction dynamics. The important role played by these bands e.g. in magnetic systems makes it relevant to seek the detailed information about the effect of disorder on their physical properties e.g if varying disorder may lead to a localization to delocalization transition  and whether its nature is similar to other disorder driven transitions.

Previous numerical studies \cite{nmg1, nmg2, cps, lbdf} on perturbed flat bands  indicate the existence of two different types of transitions: an inverse Anderson transition \cite{nmg1}, independent of disorder strength, in weak disorder regime (below a system specific disorder strength, say $w_0$) and a standard Anderson transition in strong disorder regime \cite{mj, emm}. The different nature of these transitions originates from two types of scattering mechanism prevailing in the regimes. The wavefunction interference for $w < w_0$ is caused by strong back scattering due to diverging effective mass (vanishing group velocity of the wavefunction) and is insensitive to disorder strength (disorder dependent scattering being weaker) \cite{nmg1, nmg2}. The  interference effects for $w > w_0$ are however due to disorder dominated scattering, resulting in a transition at a specific disorder if the band is single particle \cite{mj}. In case of many particle bands, the system in   $w > w_0$ regime undergoes a many body localization transition at one or more critical disorder strengths \cite{baa,mg,krav}. A theoretical formulation of the transition in weak disorder regime and its connection with the one in strong disorder regime has been missing so far.   Our objective here is to  pursue a statistical route, analyze these transitions using spectral statistics as a tool and present an exact mathematical formulation of the transition parameter in terms of the system conditions. The later helps in identifying the universality class of the spectral statistics at each type of transition and reveal analogies if any exist. 

The need to analyze the transition through  statistical approach can be explained as follows.  
The standard  search of a  localization to delocalization transition, hereafter referred as LD transition, in a disordered system is based on a range of criteria e.g. the existence of an order parameter, a divergence of correlations length at the critical point, a scaling behavior for finite system sizes and critical exponents of the average physical properties. For complex systems however the fluctuation of physical properties, from one sample to another or even within one sample subjected to a perturbation, are often comparable to their average behavior and their influence on the physical properties can not be ignored. As a consequence, one has to consider criteria based on the distribution of the physical properties \cite{mj}.  In case of systems where the physical properties can in principle be expressed in terms of the eigenvalues and eigenfunctions of a relevant linear operator, it is appropriate to seek criteria based on their joint probability distribution function (JPDF) \cite{mj}.

The definition of criticality in a JPDF of $N$ variable $x_1,\ldots, x_N$ is in general based on a single parameter scaling  concept \cite{mj}.  The distribution $P(x_1,\ldots,x_N; t_1, \ldots, t_n)$ that depends on system size $N$ and a set of $n$  parameters $t_1, t_2,\ldots, t_n$ obeys one parameter scaling if  for large $N$ it is approximately a function of only variables $x_1,..,x_N$ and one scale dependent parameter, say, $\Lambda \equiv \Lambda(N, t_1, \ldots, t_n)$. For system conditions under which the limit $\Lambda^*={\rm lim}_{N\rightarrow \infty}\;  \Lambda (N)$ exists, the distribution approaches a universal limiting form $P^*(\{x\}, \Lambda^*)= {\rm lim}_{N\rightarrow \infty} \; P(\{e\},\Lambda)$  and is referred as critical with $\Lambda^*$ as the critical parameter \cite{mj}.     
%
%
In \cite{psf1}, we considered a typical disorder perturbed flat band, with its Hamiltonian modelled by a system-dependent ensemble of Hermitian random matrices and  described a single parametric formulation of its ensemble density. As an integration of the ensemble density over all eigenfunction leads to the JPDF of its eigenvalues, this encourages us to search for a single parametric scaling of the JPDF as well as higher order eigenvalue correlations. The universal limit of these correlation, if it exists, is referred as the critical spectral statistics for the ensemble.

The concept of critical spectral statistics was first introduced in \cite{sssls} in context of metal-insulator transition in disordered Hamiltonians; the study showed that the distribution $P(s)$ of the spacings $s$ between the nearest neighbor eigenvalues of the Hamiltonian turns out to be a universal hybrid of the Wigner-Dyson distribution at small-$s$ and Poisson at large-$s$, with an exponentially decaying tail: $P(s) \sim {\rm e}^{-\kappa s}$ for $s \gg 1$ with $\kappa$ as a constant \cite{sssls}.  The analytical studies later on indicated the criticality to manifest also through an asymptotically linear behavior of the number variance $\Sigma^2(r)$ (the variance in the number of levels in an spectrum interval of length $r D$) in mean number of levels  $r$ with a fractional coefficient \cite{ckl}.

As indicated by many studies of the transition in disordered systems,  with or without particle-interactions,  the wave-functions at the critical point are multifractal \cite{mj, emm, ms}. (Note however the study \cite{gu1} claims an absence of multifractal  wavefunctions in a many body systems; see \cite{pp2} in this context). This led to introduction of the singularity spectrum as the criteria for the criticality. The wave-functions in the delocalized limit are essentially structureless and overlapping almost everywhere which leads to Wigner-Dyson type level repulsion. In localized limit, the wave-functions are typically localized at different basis state with almost negligible overlap which manifests in uncorrelated level-statistics described by Poisson universality class. But  the multifractality leads to an intimate  conspiracy between the correlations of energy levels  and eigenfunctions (for both single particle as well as many particle type).  This is because the two fractal wave-functions, irrespective of their sparsity, still overlap strongly which in turn affects the decay of level correlations at long energy ranges. For $|e_n-e_m| \gg \Delta$, the correlation between two wave-functions $\psi_n(r)$ and $\psi_m(r)$  at energy $e_n$ and $e_m$  is given as \cite{ckl}: $\langle |\psi_n(r)|^2 |\psi_m(r)|^2 \rangle  \propto |e_n-e_m|^{1-(D_2/d)}$. In \cite{ckl}, $\chi$ was suggested  to be  related to the multifractality of  eigenfunctions too: $\chi= \frac{d-D_2}{2 d}$ with $D_2$ as the fractal dimension and $d$ as the system-dimension.   However numerical studies later on indicated the result to be valid only in the weak-multifractality limit \cite{emm}.


Our objective in the present work is to analyze the criticality of the spectral statistics and eigenfunctions when a flat band is perturbed by the disorder.  In \cite{psf1},  we analyzed the disordered tight binding Hamiltonians, with at least one flat band in the clean limit, using  their matrix representation in an arbitrary basis. Presence of disorder makes it necessary to consider an ensemble of such Hamiltonians;  assuming the Gaussian disorder in on-site energies (and/or interaction strengths, hopping etc) and by representing the non-random matrix elements by a limiting Gaussian, the ensemble density, say $\rho(H)$ with $H$ as the Hamiltonian, was described in \cite{psf1} by a multi-parametric Gaussian distribution, with uncorrelated or correlated matrix elements. 
Using the complexity parameter formulation  discussed in detail in \cite{psalt, psco, psvalla, psbe, psand}, the statistics of $\rho(H)$ can then be mapped to that of a single parametric Brownian ensemble (BE) appearing between Poisson and Wigner-Dyson ensemble \cite{psbe, apbe, fh, me, fkpt5, pslg} (also equivalent to Rosenzweig-Porter model \cite{rp}). The mapping is achieved by identifying a rescaled complexity parameter of the BE with that  of the disordered band. The mapping not only implies connections of the flat band  statistics with the BE but also with other complex systems under similar global constraints e.g. symmetry conditions and conservation laws \cite{ps-all, psrp}. Additionally, as discussed in detail in \cite{psf1}, it also  leads to a single parametric formulation of the level density and inverse participation ratio of the perturbed flat band.

In case of the BEs, the existence of a critical statistics and multifractal eigenstates is already know \cite{psbe,krav}. Their connection with  disorder perturbed flat bands suggests presence of criticality 
in the latter too. 
This is indeed confirmed by our results presented here which  indicate existence of a critical statistics for all weak disorders and  is therefore in contrast to a single critical point in the disorder driven Anderson transition. Although the disorder independence of the statistics of a weakly disordered flat band was  numerically observed  in previous studies \cite{nmg2, viddi, cps}, 
 its critical aspects were not explored.   Another feature different from the Anderson transition is the following: with increasing disorder, the spectral statistics in a flat band  undergoes a Poisson $\to$  Brownian ensemble $\to$ Poisson  transition, implying a localization $\to$ extended  $\to$ localization transition of the eigenstates. As well-known, the standard Anderson transition  undergoes a delocalization $\to$ localization transition with increasing disorder \cite{psand}.   Notwithstanding these differences, the complexity parameter formulation predicts an Anderson analog of a  weakly disordered flat band and also reveals its connection of  to a wide range of other ensembles \cite{ps-all, psrp, ssps} of the same global constraint class; the prediction is verified by a numerical analysis discussed later in the paper.  Although the theoretical analysis presented here is based on the Gaussian disorder in flat bands but it  can also be extended to other type of disorders \cite{psvalla}.

The paper is organized as follows. The complexity parameter formulation for the ensemble density of a disordered tight-binding lattice, with at least one flat band in clean limit, is discussed in detail in \cite{psf1}. To avoid the repetition, we  directly proceed, in section II, to review the complexity parameter formulation for the statistics of the eigenvalues and eigenfunctions.  This formulation is used in sections III and IV to derive an exact mathematical expression for the transition parameter and seek criticality in the disorder perturbed  flat bands; here we also analyze the  influence of other neighboring bands on the statistics. A detailed numerical analysis of our theoretical claims is discussed in section V.  The next section  presents a numerical comparison of the spectral statistics of the disordered flat bands with two other disordered ensembles with dispersive bands, namely, the standard Anderson ensemble with on-site Gaussian disorder and Rosenzweig-Porter ensemble and confirms an analogy of their statistics for those system parameters which result in the same value of their complexity parameters. This in turn validates our theoretical claim regarding the existence of one parameter dependent universality class of statistics among disordered bands, irrespective of the underlying scattering mechanism, and more generally among complex systems subjected to similar global constraints e.g symmetry, conservation laws etc.  We conclude in section VII with a brief summary of our main results.

\section{ criticality of spectral statistics and eigenfunctions}

Consider the Hamiltonian $H$ of a disorder perturbed tight binding lattice with at least one flat band in clean limit: 
$H = V + U$ with $V$ and $U$ as single particle and two particle interactions. By choice of 
a physically motivated $N$-dimensional  basis,  $H$ can be represented as a $N \times N$ matrix, with $N$ as a system specific parameter \cite{oh1}.  Here we consider a basis, labelled by vectors $|k \rangle$, $k=1 \to N$,  in which (i) $H$ is Hermitian, (ii) matrix elements $H_{kl}$ are either independent or only pair-wise correlated.  (For example, for $U=0$, a basis consisting of single particle states e.g. site basis 
can serve the purpose. Similarly, for $U \not=0$ a many body wavefunction basis \cite{mg} e.g. many body Foch basis of localized single particle states or occupation number basis is appropriate \cite{oh}; see  section III of \cite{psf1} for an example). 
%
%

{\bf Ensemble complexity parameter:}
As discussed in \cite{psf1} along with a few  examples, the statistical behavior of the $H$-matrix, with  entries $H_{kl}$, can be modeled by a multi-parametric Gaussian ensemble if $H_{kl}$ are either independent or pairwise correlated.  Assuming $e_1, e_2, \ldots, e_N$ and $U_1, \ldots, U_N$ as the eigenvalues and eigenfunctions of $H$, the correlations among their various combinations can then be obtained, in principle, by an integration of the ensemble density, say $\rho(H)$, over those variables which do not appear in the combination. To study the effect of varying system conditions on the correlations, it is however easier as well as more informative to first derive an evolution equation of $\rho(H)$ which on integration leads to the evolution equations for the correlations. As described in \cite{psf1},  irrespective of the number of changing conditions, the diffusion of  $\rho(H)$ undergoes a single parametric evolution

\begin{eqnarray}
{\partial \rho\over\partial Y} &=& \sum_{k,l; q}{\partial \over 
\partial H_{kl;q}}\left[{g_{kl}\over 2}
  {\partial \over \partial H_{kl;q}} +  \gamma \; H_{kl;q}\; \right] \rho 
\label{rhoy}
\end{eqnarray}
where $g_{kl}=1+ \delta_{kl}$ with $\delta_{kl}$ as a Kronecker delta function and $\gamma$ is an arbitrary constant, marking the end of the diffusion.  The diffusion parameter $Y$, referred as the ensemble complexity parameter, is a combination of all ensemble parameters of $\rho(H)$ and thereby contains the information about the system parameters.

A detailed derivation of eq.(\ref{rhoy})  is technically complicated and  is discussed in \cite{psco}  for multi parametric Gaussian ensembles (also see \cite{psalt, ps-all}) and  in \cite{psvalla} for multi-parametric non-Gaussian ensembles. As an example,  consider the case which can be modelled by the probability density $\rho (H,v,b)=C \; {\rm exp}[{-\sum_{q=1}^\beta \sum_{k\le l} {1 \over 2 v_{kl;q}} (H_{kl;q}-b_{kl;q})^2 }]$; here $q$ refers to the real ($q=1$) or imaginary ($q=2$) component of the variable, with $\beta$ as their total number, and, the variances $v_{kl;q}$ and mean values $b_{kl;q}$ can take arbitrary values (e.g. $v_{kl;q} \to 0$ for non-random cases). Using Gaussian form of $\rho(H)$, it is easy to see that a specific combination $T \rho$ of the parametric derivatives,  namely, $T \rho \equiv \sum_{k\le l; q}\left[({2\over (2 -\delta_{kl})} \; x_{kl;q} \; {\partial \rho\over\partial v_{kl;q}} - \gamma \; b_{kl;q} \; {\partial \rho\over\partial b_{kl;q}}\right]$  can exactly be rewritten as the
right side of eq.(\ref{rhoy}) where $x_{kl;q} \equiv 1 - (2- \delta_{kl})  \gamma \; v_{kl;q}$. Clearly the left side of eq.(\ref{rhoy}) must satisfy the condition $T \rho = {\partial \rho\over\partial Y}$ which on solving gives 
$Y$ as follows \cite{psalt, psvalla}: 
\begin{eqnarray}
Y= -{1\over  \gamma \; N_{\beta}}  \; \; {\rm ln}\left[ \prod_{k \le l} \; \prod_{q=1}^{\beta}|x_{kl;q}| \quad |b_{kl;q} + b_0|^2 \right] + constant
\label{y}
\end{eqnarray}
 with $N_{\beta}={\beta N\over 2} (N+2-\beta) + N_b$ and  $N_b$ as the total number of $b_{kl;q}$ which are not zero. Further $b_0 =1$ or $0$ if $b_{kl;q}=0$ or $\not=0$ respectively.  Similarly $Y$ can be formulated for the case when  the matrix elements of $H$ are pairwise correlated; see \cite{psf1} and eq.(15) of  \cite{psco}.


{\bf Spectral density correlations: spectral complexity parameter:}
The statistical measures of a spectrum basically correspond to the  local fluctuations of spectral density around its average value and can in principle be obtained from the  $n^{th}$ order level-density correlations $R_n(e_1, e_2,..,e_n; Y)$, defined as  
$R_n= \int \prod_{k=1}^n \delta(e_k-\lambda_k) \; \rho(H; Y) \; {\rm D} H$. 
As mentioned in \cite{psf1} (see section II.C therein), eq.(\ref{rhoy}) is analogous to the Dyson's Brownian motion model of random matrix ensembles, also referred as Brownian ensemble (see section 6.13 of \cite{fh} or eq.(9.2.14) of \cite{me}). The latter describe the perturbation of a stationary Gaussian ensemble by another one with $Y$ as a perturbation parameter (or mean-square off-diagonal matrix element of the perturbation).  
Following exactly the same steps, as used in the derivation of eq.(6.14.21)  in section 6.14  of \cite{fh}, a hierarchical diffusion equation for $R_n$ can be derived by a direct integration of eq.(\ref{rhoy}) over $N-n$ eigenvalues and entire eigenvector space (also see section 8 of \cite{fkpt5} or \cite{apbe, psbe, pslg} for more information).  The specific case  of $R_1(e)$ was discussed in detail in \cite{psf1}; it varies at a scale $Y \sim N \Delta_e^2$.  The solution of the  diffusion equation for $R_2(e_1, e_2)$  with Poisson initial conditions  is discussed  in \cite{apbe} (see eq.(48) therein). Contrary to $R_1$, $R_n$ with $n >1$   undergo a rapid evolution  at a scale $Y \sim \Delta_e^2$, with $\Delta_e(e)$ as the local mean level spacing in a small energy-range around $e$. For comparison of  the local spectral fluctuations around $R_1(e)$, therefore, a rescaling (also referred as {\it unfolding}) of the eigenvalues $e_n$ by local mean level spacing $\Delta_e(e)$  is necessary. 
As discussed in detail in section 6.14 of \cite{fh} in context of single parametric Brownian ensembles, this  leads to a rescaling of both $R_n$ as well as the  crossover parameter $Y$, with new correlations given as ${\mathcal R_n}(r_1,\ldots, r_n) =\lim_{N \to \infty} \left(\Delta_e \right)^n \; R_n(e_1, e_2,..e_n)$, where $r_n = e_n /\Delta_e$ and the rescaled crossover parameter $\Lambda_e$ given as  (see eq.(6.14.12) of \cite{fh}) 

\begin{eqnarray}
\Lambda_e(Y,e)={ |Y-Y_0 | \over  \Delta_e^2}.
\label{alm1}
\end{eqnarray}

 As discussed in \cite{psalt} (see section I.E therein) and \cite{psco} (see neighborhood of eq.(53) therein), eq.(\ref{alm1}) also gives the rescaled parameter 
in context of  multi-parametric Gaussian ensembles.  (This is expected because the latter include Gaussian Brownian ensembles as a special case).  As $Y$  is a combination of all ensemble parameters, $\Lambda_e$ can be interpreted as a measure of average complexity (or uncertainty) of the system measured in units of mean level spacing. This encourages us to  refer $\Lambda_e$  as the spectral complexity parameter. It must be noted that $\Lambda_e \rightarrow \infty$ leads to a steady state i.e  Gaussian orthogonal ensemble (GOE) if $H$ is real-symmetric  ($\beta=1$) or Gaussian unitary ensemble (GUE) if $H$ is complex Hermitian  ($\beta=2$),  $\Lambda_e \rightarrow  0$   corresponds to  an initial state \cite{psalt, psco, psbe}. Also note that $\Delta_e$ here refers to the single particle mean level spacing for the single particle bands and many particle level spacing in case of the many particle bands.

In principle, all spectral fluctuation measures can be expressed in terms of ${\mathcal R_n}$; the spectral statistics as well as its criticality, therefore,  depends on the system parameters and energy only through $\Lambda_e$. For system conditions under which the limit $\Lambda^*={\rm lim}_{N\rightarrow \infty}\;  \Lambda_e (N)$ exists, ${\mathcal R_n}$ approaches a universal limiting form ${\mathcal R_n}^*(r_1,\ldots, r_n;\Lambda^*)= {\rm lim}_{N\rightarrow \infty} \; {\mathcal R}_n(r_1, \ldots, r_n;\Lambda_e)$.   Clearly the size-dependence of $\Lambda_e$ plays an important role in  locating the critical point 
which can be explained as follows. The standard definition of a phase transition refers to  infinite system sizes (i.e limit $N \to \infty$); the parameter governing the transition is therefore expected to be $N$-independent in this  limit.  In general, both $Y-Y_0$ as well as $\Delta_e$  and therefore $\Lambda_e$ can be  $N$-dependent. 
 In finite systems, a variation of $N$  therefore leads to a smooth crossover of spectral statistics  between an initial state ($\Lambda_e \rightarrow 0$) and the equilibrium ($\Lambda_e \rightarrow \infty$); the intermediate statistics belongs to an infinite family of ensembles, parametrized by $\Lambda_e$. However, for system-conditions leading to an $N$-independent value of $\Lambda_e$, say $\Lambda^*$,  the spectral statistics becomes universal for all sizes; the corresponding system conditions can then be referred as the critical conditions with $\Lambda^*$ as the critical value of $\Lambda_e$.  It should be stressed that the critical criteria  may not always be fulfilled by a given set of system conditions; the critical statistics therefore need not  be a generic  feature of all systems. (For example, it is conceivable that $\Lambda_e$ for a single particle flat band perturbed by disorder may not achieve size-independence at a specific energy for any disorder strength, thus indicating lack of criticality. Switching on particle-interactions however may change the size-dependence of $\Delta_e$ and $Y$ and lead  to a size-independent $\Lambda_e$). 
 This indicates an important application of the complexity parameter based formulation:  $\Lambda_e$ provides an exact criteria, based only on a Gaussian ensemble modeling of the Hamiltonian, to seek criticality and predict presence or absence of the LD transition in a disorder perturbed flat band (single particle as well as many particles).

At the critical value $\Lambda_e=\Lambda^*$, ${\mathcal R}_n$ (for $n >1$) and therefore all  spectral fluctuation measures  are different from the two end points of the transition i.e $\Lambda_e=0$ and $\infty$ and any one of them can, in principle, be used as a criteria for the critical statistics \cite{psbe}. 
An important aspect  of these measures is their energy-dependence: ${\mathcal R}_n$ retain the dependence through $\Lambda_e$ even after unfolding and are non-stationary  i.e vary along the spectrum \cite{psbe}. Any criteria for the  criticality in the spectral statistics can then be defined only  locally i.e within the energy range, say $\delta e_c$, in which $\Lambda_e$ is almost constant \cite{psbe}. For example,  as reported by the numerical study \cite{nmg2} of diamond lattice with two flat bands, the metal insulator transition occurs only at specific energies;  this energy dependence of transition can theoretically be explained using $\Lambda_e$ (see section IV for details).

{\bf Spectral fluctuations: standard measures:}
Based on  previous studies, numerical as well as theoretical, two spectral measures namely nearest neighbor spacing distribution $P(s)$ and the number variance $\Sigma^2(r)$ are confirmed to be a reliable criteria for seeking criticality \cite{sssls, mj, gmp, emm, me} in a wide range of complex systems. Here $P(s)$ measures the probability of a spacing $s$ between two nearest neighbor energy levels (rescaled by local mean level spacing)  and  $\Sigma^2(r)$ gives the variance of the number of levels in an interval of $r$ unit mean spacings.  
Although in past $P(s)$  has played an important role in spectral fluctuation analysis  of many body systems e.g. nuclei, atoms and molecules, the numerical rescaling of a many body spectrum is subjected to  technical issues e.g. exponentially increasing density of states or numerical simulation of large number of realization. This has motivated some recent studies to suggest another spectral measure for the short range correlations, namely, distribution of the level spacing ratio \cite{oh,bog}. In the present study, however, it is sufficient to consider $P(s)$ for the critical analysis; (this is because  the disordered systems used in our as well as previous numerical analysis \cite{nmg2} are single particle cases with Gaussian mean level densities and the unfolding on the spectrum is easier).

As confirmed by several studies in past (see for example \cite{gmp, mj, emm, me} and references therein), the level fluctuations of a system in a fully delocalized wave limit behave similar 
to that of a Wigner-Dyson ensemble i.e  GOE ($\beta=1$) for cases with time-reversal symmetry and integer angular momentum and GUE ($\beta=2$) for cases without time-reversal symmetry; here $P(s)= A_{\beta} \; s^{\beta} \; {\rm e}^{-B_{\beta} \; s^2}$ with $A_1=\pi/2, B_1=\pi/4, A_2=32/\pi^2, B_2=4/\pi$ and $\Sigma^2(r) = {2\over \pi^2 \beta} \; \left(\ln (2 \pi r) + \gamma+1+{(\beta-2)\pi^2 \over 8} \right)$ with $\gamma=0.5772$. Similarly the fully localized case shows a behavior typical of a set of uncorrelated random levels, that is, exponential decay for $P(s)$, also referred as Poisson distribution, $P(s) = {\rm e}^{-s}$, and $\Sigma^2(r) = r$ \cite{gmp,mj,me}. (In case of the structured matrices e.g. those with additional constraints besides Hermiticity however Poisson spectral statistics may appear along with delocalized eigenfunctions \cite{pstp}).  

For non-zero, finite $\Lambda_e$ cases, the exact $P(s)$ behavior is known only  for  the Brownian ensembles consisting of matrices of size $N=2$. As derived in \cite{ko}, $P(s)$  for Poisson $\to$ GOE crossover  and Poisson $\to$ GUE crossover can be given as  

\begin{eqnarray}
P(s,\Lambda_e) &=& \frac{s}{4\Lambda_e} \; {\rm exp}\left(-\frac{s^2}{8  \Lambda_e}\right) \; \int_0^{\infty} {\rm d}x \; {\rm e}^{-{x^2\over 8\Lambda_e} -x } \; I_0 \left(\frac{x s}{4\Lambda_e} \right)  \hspace{1in} \beta=1  \label{ps0} \\
P(s,\Lambda_e) &=& \frac{s}{\sqrt{2 \pi \Lambda_e}} \; {\rm exp}\left(-\frac{s^2}{8  \Lambda_e}\right) \; \int_0^{\infty} {\rm d}x \; {1\over x} \;  {\rm e}^{-{x^2\over 8\Lambda_e} -x } \; \sinh \left(\frac{x s}{4\Lambda_e} \right)  \hspace{0.5in} \beta=2 
\label{ps}
\end{eqnarray}
with $I_0$ as the modified Bessel function (see eq.(5) and eq.(11) of \cite{ko}). Here $\beta=1$ case corresponds to Brownian ensemble of real-symmetric matrices which appear as a perturbed (or non-equilibrium) state of a Poisson ensemble by a Gaussian orthogonal  ensemble (also referred as the  Poisson $\to$ GOE crossover) and are good models for systems with time-reversal symmetry. Similarly  
$\beta=2$ case corresponds to Brownian ensembles of complex Hermitian matrices,  
appearing as a perturbed state of a Poisson ensemble by a Gaussian unitary ensemble (also referred as Poisson $\to$ GUE crossover) and are applicable to systems without time-reversal symmetry. As $P(s)$ is dominated by the nearest neighbor pairs of the eigenvalues, this result is a good  approximation also for  $N \times N$ case derived in \cite{to}, especially in small-$s$ and small-$\Lambda_e$-limit. 
Using the complexity parametric based mapping of the multi-parametric Gaussian ensembles of the perturbed flat bands to Brownian ensembles, the above results can directly be used for the former case too.

As mentioned above,  $\Lambda_e$ is non-zero, finite and  size-independent in the critical regime. This along with eq.(\ref{ps0}) and eq.(\ref{ps}) indicates the following:  $P(s) \sim {\rm e}^{-\kappa s}$, for $s \gg 1$ with $\kappa$ a constant for a finite $\Lambda_e$. The study \cite{sssls} indicates an exponentially decaying tail of $P(s)$ as a criteria for critical spectral statistics. Similarly $\Sigma^2(r)$ for the critical spectral statistics is linear but with fractional coefficient: $\Sigma^2(r) \sim  \chi \; r$ with $0 < \chi <1$ \cite{mj}. The coefficient $\chi$, also referred as the level compressibility, is a characteristic of the  long-range correlations of levels; it is defined as,  in a range $r$ around energy $e$, $\chi(e, r) =1- \int_{-r}^{r} (1-R_2(e, e+s) ) \; {\rm d}s$. As $R_2(e,r)$ is related to $\Sigma_2(e,r)$,   $\chi$  can also be expressed as the $r$-rate of change of $\Sigma_2(e,r)$ \cite{mj,ckl}): $\chi = \lim_{r \to \infty}    {{\rm d} \Sigma^2(r) \over {\rm d}r}$. 
As discussed in \cite{ psand, psbe}, $\chi$ at the critical point $\Lambda^*$ can be given as 
\begin{eqnarray}
\chi &\approx & 1 - 4  \; \pi^2  \; \Lambda^*   \qquad  {\rm small} \; \Lambda^*  \label{ch1}\\
& \approx & {1\over \beta \pi^2 \Lambda^*}   \qquad  {\rm large} \; \Lambda^* 
\label{ch2}
\end{eqnarray}
with $\chi(e,r,\Lambda=0) =1$ and $0$ for Poisson and Wigner-Dyson  (GOE if $\beta=1$ or GUE if $\beta=2)$  limits, respectively.   $\chi$ is also believed to be related to the exponential decay rate of $P(s)$ for large $s$: $\chi={1\over 2 \kappa}$. Although $\chi$ is often used as a measure for criticality of the statistics \cite{mj} but, as discussed in \cite{psbe}, its numerical calculation in case of non-stationary ensembles is error-prone and unreliable.

{\bf Eigenfunction fluctuation measures:}
At the critical point, the fluctuations  of eigenvalues are in general correlated with those of  the eigenfunctions. The  spectral features  at the criticality are therefore expected to manifest in the eigenfunction measures too.  As shown by previous studies \cite{mj, emm}, this indeed occurs  through large fluctuations of their amplitudes at all length scales, and can be characterized by an infinite set of critical exponents related to the scaling of the ensemble averaged, generalized inverse participation ratio (IPR) i.e moments of the wave-function intensity  with system size. At transition, ensemble average of IPR, later defined as $ {\mathcal I}_q(e) = \int |\Psi({\bf r})|^{2q} \; {\rm d}{\bf r}$ for a state $\Psi({\bf r})$ with energy $e$, reveals an anomalous scaling with size $N$: $\langle {\mathcal I}_q \rangle(e)  \sim N^{-(q-1) D_q/d}$ with $D_q$ as the generalized fractal dimension of the wave-function structure and $d$ as the system dimension. At critical point, $D_q$ is a non-trivial function of $q$, with $0 < D_q < d$. The criticality in the eigenfunction statistics also manifests through other eigenfunction fluctuation measures e.g. IPR-distribution or two-point wave-function correlations \cite{emm}. A complexity parameter based formulation for these measures is discussed in  \cite{psbe, pslg, psf1}.

{\bf Role of dimensionality:}
The dimensionality dependence of the critical point in the localization $\to$ delocalization transitions of the wave-functions is well-established. This can also be seen through $\Lambda_e$ based formulation where dimension $d$ of the system enters mainly through local mean level spacing $\Delta_e(e)$ at energy $e$. This can be explained as follows. In the delocalized regime, a typical  state, say $\Psi({\bf r})$ occupies the  volume $L^d$ with $L$ as the linear size of the system which gives  $|\Psi({\bf r})|^2 =  {1\over L^d}$ (under normalization $\int_{L^d} |\psi({\bf r})|^2 \; {\rm d}{\bf r} =1$).  As almost all states  in this regime occupy the same space with unit probability,  $\Delta_e(e)= {1 \over {\langle \rho_e \rangle} \; L^d}$ with ${\langle \rho_e(e) \rangle}$ as the mean spectral density (i.e number of states per unit energy per unit volume): ${\langle \rho_e(e) \rangle}= {1\over N} \langle \sum_{n=1}^N \delta(e-e_n) \rangle = {R_1 \over N}$.  In the localized regime, the states are typically not overlapping but localized in the same regime with a  probability ${\xi^d \over L^d}$ where $\xi$ is the average localization length  at  energy $e$;  consequently $\Delta_e(e)$ in this case corresponds to the level spacing in the localized volume $\xi^d$ and is given as $\Delta_e(e)={1\over \langle \rho_e \rangle \; \xi^d}= {N \over R_1 \; \xi^d}$. 
%
%
Note $\xi(e)$  is in general a function of dimensionality  \cite{mj} (besides other system conditions e.g. particle interactions) and can be expressed in terms of  the  inverse participation ratio $\overline{\langle {\mathcal I}_2 \rangle}$ of the eigenfunctions in a small neighborhood of $e$ (with $\langle . \rangle$ and $\overline{.}$ implying ensemble and spectral averages respectively): $\xi^d =(\overline{\langle {\mathcal I}_2 \rangle})^{-1}$. The above gives 
$\Delta_e(e)= {N  \over   R_1} \; \overline{\langle {\mathcal I}_2 \rangle}$ which on substitution in eq.(\ref{alm1})   results in  

\begin{eqnarray}
\Lambda_e(Y,N, e)  = {|Y-Y_0 | \over  N^2} \left({R_1 \over \overline{\langle {\mathcal I}_2 \rangle}}\right)^{2}.
\label{almx}
\end{eqnarray}
As clear from the above, a size-independence of $\Lambda_e(e)$ i.e existence of $\Lambda^*(e)$ requires a subtle cancellation of size-dependence among the ensemble complexity parameter $Y$, ensemble averaged level density $R_1$ and inverse participation ratio ${\mathcal I}_2$ (single particle or many particle based on the nature of the band). 
Note, in case of a many particle band, $\xi$ refers to many particle localization length, defined as the typical scale at which many-particle wavefunction decays and $I_2$ its inverse participation ratio.

In the following sections, we use eq.(\ref{almx}) to derive $\Lambda_e$ for three cases of disorder perturbed flat bands;  $R_1$ and $I_2$ for these cases are derived in \cite{psf1}.

\section{Transition in an isolated flat band}

In \cite{psf1}, we obtained the ensemble complexity parameter $Y$ for a perturbed  flat band.
For cases, in which disorder $w$ is the only parameter subjected to variation, $Y$ turns out to be

\begin{eqnarray}
Y-Y_0 =- {1\over  N} \; {\ln|1- w^2|}, 
\label{yw}
\end{eqnarray} 
where $Y_0$ corresponds to the unperturbed flat band  ($w=0$) and $N$ is the number of energy levels in the band.

As discussed in \cite{psf1}, the level density $R_1$ for an isolated flat band for arbitrary $w$ is (eq.(39) of \cite{psf1})

\begin{eqnarray}
R_1(e; w) =   {N\over \sqrt{2 \pi w^2}} \; {\rm e}^{-{e^2 \over 2 w^2 }}
\label{g5}
\end{eqnarray}

Further the averaged inverse participation ratio ${\overline {\langle  {\mathcal I}_2  \rangle}}(e)$ for arbitrary $w$ and large $N$ can be approximated as (see section V.B of \cite{psf1})

 \begin{eqnarray}
{\overline {\langle  {\mathcal I}_2  \rangle}}  & \approx &  
{6 \; \pi \; u_0 \over N \; E_c} \;  {\rm e}^{{2 \Lambda_I \over N}} 
 {\rm e}^{-{4 e \over E_c} + {e^2 \over 2 w^2}} 
\label{aq9}
 \end{eqnarray}

with $u_0$ as the local intensity at $e=0$ and $\Lambda_I= {4 \ln|1- w^2| \over E_c^2}$. Here $E_c$ is an energy scale associated with the range of level-repulsion around $e$ and can in general depend on $e$ as well as $w$. 
Eq.(\ref{aq9}) is obtained by assuming   $E_c \sim  N^{-\mu}$ with $\mu \ge 0$ which is consistent with the definition of $E_c$;  
as discussed in \cite{psf1}, $E_c \sim E_{th}$ with $E_{th}$ as the Thouless energy: $E_{th} \sim o(N^{-1})$ and $o(N^0)$ for the localized and delocalized dynamics respectively but in partially localized regime $E_{th} \sim \Delta(e). N^{D_2/d}$, with $\Delta(e) = (R_1(e))^{-1}$ as the mean level spacing at energy $e$, $D_2$ as the fractal dimension and $d$ as the physical dimension. Assuming $\Delta(e)   \sim N^{-\eta}$ with $\eta$ as a system-dependent power, this gives 
\begin{eqnarray}
E_c \sim N^{-(\eta d-D_2)/d}
\label{ec}
\end{eqnarray}
 and $\mu=(\eta d -D_2)/d$. 
With $0 \le D_2 \le d$, the assumption $\mu >0$ is valid at least in flat band regime where $\eta =1$ (the latter follows from eq.(\ref{g5})).

Substitution of eq.(\ref{g5}), eq.(\ref{aq9})  along with eq.(\ref{yw}) in eq.(\ref{almx})  leads to 

\begin{eqnarray}
\Lambda_e(Y,N, e)  
=  {N E_c^2  \over 72 \; \pi^3 \; u_0^2} \; {|\ln|1-w^2| |\over w^2} \;
 {\rm e}^{-{16 \ln|1-w^2| \over N E_c^2}} \;  {\rm e}^{{8 e \over E_c} - {2 e^2 \over  w^2}} 
\label{alm3}
\end{eqnarray}
As clear from the above, $\Lambda_e$ depends on the energy $e$, disorder $w$ as well as energy scale $E_c$. To seek the critical point, it is necessary to find specific $e$ and $w$ values which results in  a  $\Lambda_e$ 
size-independent as well as different from the two end-points: $ {\lim \atop N \to \infty} \; \Lambda_e  \not= 0, \infty  $. For further analysis of eq.(\ref{alm3}), we consider following energy and disorder regimes:

{\it Case $e \sim 0$:} For large $N$ and $E_c \sim N^{-\mu}$ with  $0 <\mu \le 1/2$, one can approximate ${\rm e}^{-{16 \ln|1-w^2| \over N E_c^2}} \sim 1$. This along with eq.(\ref{alm3}) then implies  disorder-independence of $\Lambda_e$ for $e\sqrt{2} < w < 1$:
$\Lambda_e(Y,N, e)  = {N E_c^2  \over 72 \; \pi^3 \; u_0^2} $. Further for cases with $\mu =1/2$, $\Lambda_e$ is also size-independent, implying a critical spectral statistics in the bulk of the flat band spectrum (i.e $e \sim 0$).   As indicated by our numerical analysis, $\eta=1, D_2 \approx 1.18$ which gives  $E_c \sim N^{-0.41}$ for the two dimensional chequered board lattice ($d=2$) in weak disorder limit.  The criticality of the spectral statistics is also confirmed by the size-independence of the fluctuation measures  (see parts (c) and (e)  of the figures (2,3)). The details are discussed later in section V.  (Note, for weak disorder, the chequered board lattice has a perturbed flat band in the neighborhood of a dispersive band but the former can still be treated as isolated).

%

For large $w$ and finite $N$, $\Lambda_e$ decrease smoothly with increasing $w$ and therefore the spectral statistics near $e \sim 0$ again approaches Poisson limit, implying lack of level-repulsion.     Further in limit $ N \to \infty$, $\Lambda_e \to 0$ for any finite $w >1$ which indicates a transition from critical statistics to Poisson at $w \approx 1$. As clear from the above, the statistics undergoes an inverse Anderson transition 
in the disorder perturbed flat band, with fully localized states at zero disorder becoming partially localized for a  weak disorder ($w < 1$ in our case). However the usual Anderson transition sets in presence of  strong disorder ($w \simeq 1$). In infinite size limit $N \to \infty$, the statistics therefore shows two types of disorder driven critical behavior near $e \sim 0$: 
(i) at $w \sim 0$, Poisson $\rightarrow$ near GOE (or near GUE in presence of magnetic field) transition of the level statistics,
(ii) at $w \sim 1$, the level-statistics transits from near GOE/ GUE $\rightarrow$ Poisson.  


{\it Case $e >0$:}  For $w^2 < 2 e^2$, the term ${\rm e}^{- {2 e^2 \over  w^2}} \to 0$ which gives $\Lambda_e \to 0$ and Poisson statistics. But, for a fixed $e >0$, ${\rm e}^{- {2 e^2 \over  w^2}} \to 1$ with  increasing $w$ and consequently $\Lambda_e$ increases too if $w < 1$. For $w >1$, however, the contribution from other terms results in a decrease of $\Lambda_e$ with increasing $w$. For finite $N$ the statistics at $e >0$ therefore changes  from Poisson $\to$ GOE $\to$ Poisson with increasing $w$.  




An important point worth emphasizing here is an energy dependence of the spectral statistics for infinite system sizes ($N \to \infty$) and for weak disorder: critical near $e \sim 0$ if $E_c(e \sim 0) \propto {1\over \sqrt{N}}$  but Poisson for $e>0$ if $N \; E_c^2 \le 1$ for $ e>0$. This suggests the existence of a mobility edge separating partially localized states from the localized states.

At this stage, it is relevant to indicate the following.  
As the level density for a flat band in clean limit can be expressed as a $\delta$-function, irrespective of whether the band is single or many particle type, the formulation derived in \cite{psf1} remains valid for both type of bands; (although $Y$ for two cases is different). Similarly the response of the average inverse participation ratio to weak disorder discussed in \cite{psf1} is based on a knowledge of initial condition only and not on the presence or absence of interactions in the band; it is thus applicable for both type of bands too. This is however not the case for the spectral fluctuations which are governed by $\Lambda_e$ and therefore dependent on the local mean level spacing $\Delta_e$. For many particle spectrum, $\Delta_e$ in general depends on many particle  localization length which can be varied by tuning either disorder or interactions. Thus the size-independence of many body $\Delta_e$ can  be achieved in many ways which could as a result lead to  more than one critical point.

\section{Transition in a flat band with other bands in the neighborhood}


In presence of other bands, the energy as well as size dependence of $\Lambda_e$, defined in eq.(\ref{almx}) can vary significantly based on the neighborhood. 
As calculation of $\Lambda_e$ requires a prior knowledge of  the level densities and IPR, 
here we consider two examples for which these measures are discussed in \cite{psf1}:

(i)  {\it two flat bands}: 

As discussed in section VI.A  \cite{psf1}, $R_1(e)$  can now be expressed as a sum over two Gaussians (originating from $\delta$-function densities of two flat bands)

\begin{eqnarray}
R_1(e; w) =   {N\over 2 \; \sqrt{2 \pi w^2}} \; \sum_{k=1}^2 {\rm e}^{-{(e-e_k)^2 \over 2 w^2} }  
\label{r2}
\end{eqnarray}
with $e_1, e_2$ as the centers of two flat bands. The IPR  in large $N$ limit is (see section VI.B of \cite{psf1})

\begin{eqnarray}
{\overline {\langle  {\mathcal I}_2  \rangle}}(e,\Lambda_I)  & \approx &  
 { 3 \; u_0  \sqrt{2 \pi} \over 2 \; R_1 \; E_c}  \;
 \sum_{k,l=1}^2 {\rm e}^{-{4 (e-e_k) \over E_c} } \;  {\rm e}^{-{ (e_l-e_k)^2 \over 2 w^2}+ {2\Lambda_I \over N}}  \; \Theta(e-e_k)
\label{i2}
\end{eqnarray}
with $\Lambda_I={4 \ln|1-w^2| \over E_c^2}$ and
$\Theta(x)$ as the step function: $\Theta(x) =0,1$ for $x <0$ and $x > 0$ respectively.
Substitution of eq.(\ref{r2}) and eq.(\ref{i2}) in eq.(\ref{almx}) now gives $\Lambda_e$ for this case. A better insight can however be gained by deriving $\Lambda_e$ in different energy regimes.

{\it Case $e \sim e_k$:}
   For $e \sim e_k$, with $k=1,2$, eq.(\ref{r2}) and eq.(\ref{i2}) can be approximated as 
$R_1(e; w) \approx   {N\over 2 \sqrt{2 \pi w^2}}[1+  {\rm e}^{-{ (e_2-e_1)^2 \over 2 w^2}}]$  
and ${\overline {\langle  {\mathcal I}_2  \rangle}} (e)   
\approx   {6 \; \pi \; u_0 \over N \; E_c} \;  
 {\rm e}^{{8 ln|1-w^2| \over N E_c^2}}$. These on substitution in eq.(\ref{almx}) give
\begin{eqnarray}
\Lambda_e(Y,N, e)  
\approx  {N E_c^2  \over 288 \; \pi^3 \; u_0^2 }  \;  {|\ln |1-w^2| |\over w^2}  \; {{\rm e}^{-{16 \ln|1-w^2| \over N E_c^2}} \over [1+  {\rm e}^{-{ (e_2-e_1)^2 \over 2 w^2}}]^2}
\label{alm4}
\end{eqnarray}
Clearly, similar to the single band case, here again $\Lambda_e$ is independent of disorder  for $w < 1$
 and in large $N$ limit but it rapidly decreases with larger disorder (for $w >1$).  Here again the  size-independence of $\Lambda_e$ requires $E_c \propto {1\over \sqrt{N}}$.  
For $w <1$,  the spectral statistics at the centers of two Gaussian bands (flat ands in clean limit)
 can therefore be critical as well as disorder independent only if  $\mu=1/2$. 

{\it Case $e \sim (e_1+e_2)/2$:}
For the energies  midway between two bands,   $R_1$ is very small for $w <1$ but, contrary to band center, it increases with increasing $w$ for $w > |e_1-e_2|$: $R_1\left({e_1+e_2\over 2}\right) = { N\over \sqrt{2 \pi w^2}} \;  {\rm e}^{-{(e_2-e_1)^2 \over 8 w^2} }$  
and eq.(\ref{i2}) gives
${\overline {\langle  {\mathcal I}_2  \rangle}}(e) 
\approx    {6\;\pi \; u_0 \over N \; E_c} \;  
 {\rm e}^{{8 \; \ln |1-w^2| \over N E_c^2}}  \; {\rm e}^{-2 (e_2-e_1) \over E_c} \;  {\rm e}^{(e_1-e_2)^2 \over 8 w^2} \; \left[ 1 +  {\rm e}^{-(e_1-e_2)^2 \over 2 w^2} \right]$. With $Y-Y_0$ given by eq.(\ref{yw}), we now have 
\begin{eqnarray}
\Lambda_e(Y,N, e)  
=  { N E_c^2  \over 72 \; \pi^3 \; u_0^2}  \;  {|\ln |1-w^2| |\over w^2}  \; {\rm e}^{-{16 \;\ln |1-w^2| \over N E_c^2}} \; \;{ {\rm e}^{4 (e_2-e_1) \over E_c} \; {\rm e}^{- (e_1-e_2)^2 \over 2 w^2} \; \over \left(1 +  {\rm e}^{-(e_1-e_2)^2 \over 2 w^2} \right)^2}
\label{alm5}
\end{eqnarray}
As clear from the above, here also $\Lambda_e$ become $N$-independent thus implying critical statistics  if $E_c \propto N^{-1/2}$.  Note however the term ${\rm e}^{- (e_1-e_2)^2 \over 2 w^2}$ present in eq.(\ref{alm5}) can result in the statistics different  from that of $e \sim e_k$.

A case of two flat bands was studied in \cite{nmg2} for the 3-dimensional hexagonal diamond lattice. The study indicates $D_2 \approx 2.55$ and $2.61$ for $e \sim e_k$ and $e\sim (e_1+e_2)/2$, respectively. With $\Delta(e) \propto N^{-1}$ and $d=3$, eq.(\ref{ec}) gives  $E_c$ for this system as $N^{-0.15}$ for $e \sim e_k$ and $N^{-0.13}$  for $e\sim (e_1+e_2)/2$. Based on our theory, the statistics is predicted to be size as well as disorder dependent near $e \sim (e_1+e_2)/2$ and size-dependent but disorder-independent near $e \sim e_k$. The display in figures (4,5) of \cite{nmg2} indeed confirms this prediction.

The case of three flat bands was discussed in \cite{vidi}, for a bipartite periodic lattice described by a tight binding, interacting  Hamiltonian. The study indicates a localization $\to$ delocalization transition at the onset of disorder or many body interactions.  The possibility of a critical behavior for this case can be explored along the same route as given above.

(iii)  {\it a flat band at the edge of a dispersive band}: For the combination of a flat band located at $e=0$ and a dispersive band  at $ e >0$ with the level density $f_d(e)$, the results in section VI of  \cite{psf1} give 

\begin{eqnarray}
R_1(e; w) =   {N\over 2 \; \sqrt{2 \pi w^2}} \;{\rm e}^{-{e^2 \over 2 w^2} }  + {N\over 2} \; f_w(e, w, N)
\label{gf4r}
\end{eqnarray}

with $f_w(e,w,N)$ as the dispersive band density at disorder $w$
and 

\begin{eqnarray}
 {\overline {\langle  {\mathcal I}_2  \rangle}}(e,\Lambda_I)  & \approx &    {1\over 2 \; R_1} \;
{3  \sqrt{2} \over w \; E_c}\;\left[ u_0  \sqrt{\pi}  +B_1+B_2 + B_3 \right] \;  {\rm e}^{-{4 e \over E_c} + {2 \Lambda_I \over N}}   
\label{iq9}
\end{eqnarray}

with 
 $B_1 = {2 u_0 w \over E_c} \; \sqrt{\pi N\over \Lambda_I} \; \int_{-\infty}^{\infty} {\rm d}x \; f_w(x)  \; {\rm e}^{-{2 N x^2\over \Lambda_I E_c^2} +{4 x \over E_c}}$,
$B_2 = N \int_{-\infty}^{\infty} {\rm d}x \; f_w(x) \; u_d(x) \; {\rm e}^{-{x^2\over 2 w^2} +{4 x \over E_c}}$,
$B_3 = {\sqrt{2 \pi w^2} } \; \int_{-\infty}^{\infty} {\rm d}x \; f_w(x) \; u_d(x) \; 
{\rm e}^{4 x \over E_c}$ and $\Lambda_I = {4 \ln|1-w^2|\over E_c^2}$.
Here  $u_0$ and $u_d(e,w)$ are the local eigenfunction intensities in the flat band at disorder $w=0$ and in dispersive band at disorder $w$.   For cases in which $f_w(e, w,N)$ varying slower than the Gaussians in the related integrals,  $B_1$ and $B_2$ can be  approximated as follows: 
$B_1= \pi \sqrt{2} u_0 \; f_w \left({\Lambda_I E_c \over 4}\right) \; {\rm e}^{2 \Lambda_N \over N}$, 
$B_2=\sqrt{2 \pi w^2} \; u_d \left({4 w^2 \over E_c^2} \right) \; {\rm e}^{8 w^2 \over E_c^2}$.

A substitution of eq.(\ref{gf4r}), eq.(\ref{iq9}) along with eq.(\ref{yw}) in eq.(\ref{almx}) give $\Lambda_e$ for arbitrary energy and disorder but here again it is instructive to analyze the behavior near 
specific energies:

{\it Case $e \sim 0$:}
Due to almost negligible contribution for weak disorder from the dispersive part near $e \sim 0$, 
one can approximate $R_1 \approx {N \over 2 \; \sqrt{2 \pi w^2}}$ and ${\overline {\langle  {\mathcal I}_2  \rangle}} \approx {6 {\pi} u_0\over N E_c} $  which in turn gives $\Lambda_e={N \; E_c^2 \over 288 \pi^3 u_0^2}$. The latter is therefore again size as well as disorder independent indicating criticality near $e \sim 0$ for all weak-disorders if $E_c \propto N^{-1/2}$. As intuitively expected, the behavior of spectral statistics near $e \sim 0$ and $w <1$ in this case is analogous to that of the single flat band case. 

As mentioned in \cite{cps, psf1}, the two dimensional chequered board lattice consists of a flat band and a dispersive band in clean limit. 
Our numerical analysis of the system for $w <1$ indicated   $\Delta(e) \propto N^{-1}$ and $D_2 \sim 1.18$ (see figures 2(a,b), 3(a,b) of the present work and figure 4 of \cite{psf1}), leading to $E_c \sim N^{-0.41}$  which implies ${\overline {\langle  {\mathcal I}_2  \rangle}} \sim N^{-0.59}$, an indicator of partially localized states \cite{pp3}.
Based on theoretical grounds, therefore, the  spectral statistics is expected to be critical near $ e \sim 0$ and $w <1$; this is indeed confirmed by the size-independence of the statistics  displayed in figure 2(c,e) and figure 3(c,e).


For large $w$ (e.g $w >1$ for the case with $\mu=1/2$), however the contribution from the dispersive band becomes significant near $e \sim 0$. This results in 
$R_1(e \sim 0) \approx {N \over 2 \; \sqrt{2 \pi w^2}}  \; T_1$ where $T_1 = 1+ \sqrt{2 \pi w^2} \; f_w(0,w,N)$ and  
${\overline {\langle  {\mathcal I}_2  \rangle}}  \approx   {6\;  \sqrt{\pi} \over N \; E_c \; T_1} \;  \left[ u_0 \sqrt{\pi} + B_1+B_2 + B_3\right] \;  {\rm e}^{ {8  \ln |1-w^2| \over N E_c^2}}$. 
These on substitution in eq.(\ref{almx}) give 
%
%

\begin{eqnarray}
\Lambda_e(Y,N, e)  
=  {N E_c^2  \over 288 \; \pi^2 }  \;  \;  {|\ln |1-w^2| |\over w^2}  \;  {T_1^4 \over \left(u_0 \sqrt{\pi} + B_1+ B_2 + B_3 \right)^2}  \;  {\rm e}^{-{16 \;\ln|1- w^2| \over N E_c^2}} 
\label{alm6}
\end{eqnarray}

As clear from the above, for large $w$ and finite $N$, $\Lambda_e$ decrease smoothly with increasing $w$ and therefore the spectral statistics near $e \sim 0$ again approaches Poisson limit, implying lack of level-repulsion; note $E_c$ is expected to decrease with increasing $w$. 
 Further in limit $ N \to \infty$, $\Lambda_e \to 0$ for any finite $w >1$ which indicates a transition from critical statistics to Poisson at $w \approx 1$. 

 The above prediction is again  consistent with our numerical analysis (see figure 4(c,e) and figure 5(c,e)). Note as displayed in figure 5(a), $\Delta(e) \propto N^{-1}$ and figure 5(b), $D_2 \approx 0.5$ which gives $E_c \sim N^{-0.75}$, thus implying a size-dependent $\Lambda_e$, approaching zero in large $N$-limit which corresponds 
 to Poisson statistics. Figures 4(c,e) and  figure 5(c,e) indeed confirm the approach of spectral measures to  Poisson limit for $e \sim 0$ and $w >1$.  
 

{\it Case $e > 0$:}  Due to weaker contribution from the Gaussian density for $ e >0$, the contribution from the dispersive band density need not be negligible and  it is appropriate to consider the full form of $R_1(e)$.   The IPR can now be approximated as 

\begin{eqnarray}
{\overline {\langle  {\mathcal I}_2  \rangle}}  \approx   {6\; \sqrt{\pi} \over N \; E_c \; T_0} 
 \; \left[ u_0 \sqrt{\pi} + B_1+B_2 + B_3\right] \; {\rm e}^{-{4 e \over E_c}  + {8 \ln |1-w^2| \over N E_c^2}} 
\label{iq15}
\end{eqnarray}
where $T_0 = {\rm e}^{-{e^2 \over 2 w^2}}+\sqrt{2 \pi w^2}\; f_w(e) $.
The above leads to 

\begin{eqnarray}
 \Lambda_e(Y,N, e)  
=  {N E_c^2  \over  288 \; \pi^2 }  \;  \;  {|\ln |1-w^2| |\over w^2}  \;  {T_0^4 \over \left(u_0 \sqrt{\pi} + B_1+ B_2 + B_3 \right)^2}   \;  {\rm e}^{-{16 \ln |1-w^2| \over N E_c^2} } \; 
{\rm e}^{{8 e \over E_c}} 
\label{alm7}
\end{eqnarray}

The presence of term ${\rm e}^{{8 e \over E_c}} $ in eq.(\ref{alm7}) results in the statistics different from the case $ e \sim 0$. For $w=0$, the statistics in the dispersive band at $e >0$ is that of a GOE  (or GUE if time-reversal symmetry is violated) but, with onset of disorder, it abruptly changes to Poisson.
With increasing $w$ for $0 < w <1$, $\Lambda_e$ increases but starts decreasing above $w =1$. For large $e>0$, the statistics therefore varies from GOE (at $w=0$) to Poisson statistics for $w = 0^+$, becomes GOE at $w=1$,  and then   again approaches Poisson $w >1$. This prediction is consistent with our numerical results displayed in figures 2(d,f), 3(d,f) for $w <1$ and figures 4(d,f) and 5(d,f) for $w \ge 1$.



\section{Numerical analysis: 2-d chequered Board Lattice}

To verify our theoretical predictions, we pursue a numerical statistical analysis of the eigenvalues and eigenfunctions of  the Hamiltonian 
$H=\sum_{x,y}^N  V_{xy} \;  c_y^{\dagger}. c_x$ of a 2-$d$-planer pyrochlore lattice with single orbital per site \cite{cps, psf1}. With 2-d unit cell labeled as $(m,n)$, one can write a site-index as $x=(m,n,\alpha)$ with $\alpha=a,b$ (i.e two atoms per unit cell). The lattice consists of one flat band   $ E_f = \varepsilon - 2 t$ and one dispersive band  $E_d=\varepsilon+2t (\cos k_x + \cos k_y +1)$ if $V_{xy}$ satisfies following set of conditions \cite{psf1, cps}:
(i) $V_{xx}=\epsilon$, (ii) $V_{xy}=t$ with $x=(m,n,\alpha)$ if $y=(m,n,\beta)$ or $(m-1,n,\beta)$ or $(m,n+1, \beta)$ with $\beta=a,b$ and (iii) $V_{xy}=0$ for all other $x,y$ pairs. 

For $\epsilon=2, t=1$, the Hamiltonian, in absence of disorder, consists of a flat band at $e =0$ and a dispersive band centered at $e=4$.   (This can be seen  from the band energies $E_f$ and $E_d$ given above).  
The onset of disorder through on-site energies with $\langle V_{xx} \rangle=\varepsilon, \langle V_{xx}^2 \rangle - \langle V_{xx} \rangle^2 = w^2$  leads to randomization of the Hamiltonian. For the numerical analysis, therefore,  we simulate large matrix ensembles of the Hamiltonian, and  at many $w$, for various ensemble-sizes $M$ (the number 
of matrices in the ensemble) as well as the  matrix-sizes $N=L^2$. The energy-sensitivity of the transition (due to energy-dependence of $\Lambda_e$) requires the fluctuations analysis at precisely a given value 
of energy.  In order to improve the statistics however a consideration of the averages over an optimized energy range  $\Delta E$ is necessary (not too large, to avoid  
mixing of different statistics).  For comparison of a measure for different system-sizes $N$ at a given disorder,   we have used only $20 \%$ levels in our numerical analysis.
 

In \cite{psf1}, we theoretically analyzed the disorder dependence of level density $R_1$ and average inverse participation ratio $\langle I_2 \rangle$. Our results indicated a disorder insensitivity of these measures in weak disorder limit ($w <1$).  This was also confirmed by their numerical analysis as well as that of $D_q$  displayed in figure 4 of \cite{psf1}. A search for criticality however also requires an analysis of the size-dependence of the fluctuation measures.
In this section, we numerically analyze the disorder and size dependence of the spectral fluctuations as well as the fractal dimensions $D_q$. The figure 1  displays the disorder-dependence of $P(s)$ and $\Sigma^2(r)$ in two energy regimes i.e near $e \sim 0$ and $e \sim 4$  (corresponding to bulk of the flat band and dispersive bands in clean limit).
As clear from figures 1(a) and 1(c), for a weak disorder ($w <1$) and near $e \sim 0$, both measures are insensitive to change in disorder.  But as displayed in figures 1(b,d), the statistics in the dispersive band ($e \sim 4$)  varies with disorder even for weak disorders. A similar result was reported by the numerical study of a $3$-dimensional disordered diamond lattice (with two flat bands in the clean limit) \cite{nmg2}.    The effect of on-site disorder for  the  ${\mathcal T}_3$ lattice with three flat bands in clean limit, was analyzed in \cite{viddi}. The results again indicated disorder independence of the  fluctuation measures for low disorder $w <1$ but an increase of localization with $w$ for 
$w >1$.

Our next step is to seek criticality in the spectral and eigenfunction statistics. For this purpose, we focus on the  size-dependence of $P(S), \chi$ and $D_2$ in two energy  regimes $e \sim 0$ and $e \sim 4$;  the results for four disorder-strengths, two in weak and two in strong disorder regime, are displayed in figures 2-5. (Here, for clarity of presentation, a comparison with theoretical approximation given by eq.(\ref{ps0}) is not displayed). To determine $\Lambda_e$ for these cases, it is numerically easier to use the following expression 
(instead of the theoretical approximation discussed in the previous section),
\begin{eqnarray}
\Lambda_{e, FE} =  {R_1^2 \over \langle I_2 \rangle^2} \; {|\ln |1-w^2| |\over  N^3} .
\label{alm10}
\end{eqnarray}
where $R_1$ and $\langle I_2 \rangle$ are numerically obtained; the corresponding values are given in the captions of figures 2-5.   Before proceeding further, it is important to note that the intial condition $w=0$ (clean limit) corresponds to $\Lambda_e=0$ but the initial state of the statistics is different in the two bands. In clean limit, the flat band  corresponds to Poisson statistics while dispersive band corresponds to that of the GOE.

The size-independence as well as location of the curves, intermediate to Poisson and GOE limits in figure 2(c,e) is an indicator of the critical spectral statistics; note the disorder here is very weak ($w \sim 10^{-5}$). Similarly behavior in figure 2(b) is an indicator of the partially localized wave-functions \cite{pp3, emm} in the weakly disordered flat band bulk; also note that figures 2(b) and 2(e) give $D_2 \approx 1.2$ and $\chi \approx 0.2$ respectively for the flat band which agrees well with the prediction based on the weak multifractality relation  $D_2=d(1-2 \chi)$  (note $d=2$ in our case) \cite{ckl}.  With $\Lambda_e \approx 0.384$ in this case (see caption of figure 2), the numerically obtained $\chi$-value is also consistent with eq.(\ref{ch2}). In contrast to behavior near $e\sim 0$, the size-dependence of the measures  is clearly visible from figures 2(d,f) (depicting behavior near $e \sim 4$) which rules out criticality in the dispersive regime. Furthermore the statistics here is almost Poisson which indicates an abrupt transition from GOE (for $w=0$) with onset of disorder; 


As shown in figures 3(b,c,e), the critical behavior in the flat band persists even when disorder is varied to $w \sim 10^{-1}$. But in contrast to $w \sim 10^{-5}$,  the statistics in the dispersive regime ($e \sim 4$) now shifts away from the Poisson limit (see figures 2(d,f) and 3(d,f)); this implies a tendency of the wave-functions in the dispersive band to increasingly delocalize as $w$ approaches $1$.  The results given in figures 2 and 3 clearly indicate the reverse trend of the statistics in two bands with increasing disorder in range $0 < w \le 1$: the flat band bulk undergoes a Poisson $\to$ near GOE $\to$ near Poisson type crossover with increasing $w$ (though never reaching GOE) but the dispersive bulk changes from  GOE $\to$  Poisson $\to$  GOE limit. For $w >1$ however bands increasingly overlap with each other and the statistics for both energy ranges approaches  Poisson limit with increasing disorder (although at different rate based on energy regime, see figures 4,5), as expected from a standard Anderson transition (later discussed in more detail in \cite{emm}). The statistics now seems to be size-independent for all energy ranges. Also note from figures 5(b,e), the relation $D_2 =d(1-2 \chi)$ is no longer so well-satisfied near $e \sim 0$ 
 (here $d=2$, $D_2 \approx 0.5$ from figure 5(b) and $\chi \approx 0.42$ from figure 5(e)). This is expected because the multifractality in the band is no longer weak.

As confirmed by a large number of theoretical, numerical as well as experimental studies of wide-ranging complex systems \cite{gmp, mj, emm, me},  Poisson and GOE type behavior of the spectral statistics are indicators of localized and delocalized dynamics of the eigenfunctions, respectively, with an intermediate statistics indicating  partially localized states \cite{pp3}; (note, as discussed in \cite{pstp}, the above relation between spectral statistics and eigenfunction dynamics is valid only for Hermitian matrices). This implies that, for $w \sim 10^{-5}$ and $10^{-1}$, the states near $e \sim 0$ are extended (although not completely delocalized) but localized near $e \sim 4$ (see parts (c),(d) of figures 2,3). For  $w=1$, however the localization tendency is now reversed, with almost localized states near $e \sim 0$ but delocalized near $e \sim 4$. This inverse eigenstate localization tendency at $e \sim 0$ to the at $e \sim 4$ for a given weak disorder hints at the  existence of a mobility edge/region. Note beyond $w >1$, all states are almost localized although the rate of change of localization length with disorder strength is energy-dependent (This follows because the average localization length in general depends on both disorder as well as energy).  

Let us now focus on the flat band only. 
As clear from the above, the behavior near $e \sim 0$ indicates the occurrence of an inverse Anderson transition, with fully/ compact localized  states at zero disorder becoming partially localized for a non-zero weak disorder ($w < 1$ in our case). However the usual Anderson transition sets in presence of the strong disorder (for $w \ge 1$). The quantum dynamics near $e \sim 0$ now shows two types of critical behavior: 
(i) at $w=0$, a localized $\to$ extended state transition , in weak disorder regime and
(ii) an extended state  $\to$ localization transition at $w \approx 1$.

\section{Analogy with other ensembles}

Based on the complexity parametric formulation,  different ensembles subjected to same global constraint (which  is the  Hermitian nature of $H$-matrix in present study) are expected to undergo similar evolution. This in turn implies an analogy of their statistical measures if the values of their complexity parameters are equal and the initial conditions are statistically analogous. In this section, we verify the analogy by  comparing the statistical behavior of weakly disordered flat bands with two other disordered ensembles of real-symmetric matrices, namely, the Anderson ensemble with on-site Gaussian disorder and Rosenzweig-Porter ensemble.  Similar to flat band lattices, both of these ensembles can be expressed  as a multi-parametric Gaussian ensemble and the expressions for $Y$ and $\Lambda_e$ for them can be easily obtained (see \cite{psand}, \cite{psrp} and \cite{psbe} for details).  The two ensembles can briefly be described as follows.

\vspace{0.1in}

{\bf Anderson Ensemble:} 
The standard Anderson Hamiltonian $H= \sum_{k=1}^N  \varepsilon_k \; c_k^{\dagger}. c_k +   
 \sum_{k,l=n.n}^N  V_{kl} \;  c_k^{\dagger} . c_l$ describes the dynamics of an electron moving in a random potential in a $d$-dimensional tight binding lattice with one atom per unit cell. The disorder in the lattice can appear through on-site energies $\varepsilon_k$ or  hopping $V_{kl}$ between nearest neighbor sites. Here we consider the lattice with $N$ sites, an on-site Gaussian disorder (with $\langle \varepsilon_k^2 \rangle = w^2$, $\langle \varepsilon_k \rangle=0$,  and a random nearest neighbor hopping ($\langle V_{kl}^2 \rangle=t \; f_0, \langle V_{kl} \rangle=0$ with $f_0 =1$ if the sites $k,l$ are nearest neighbors otherwise it is zero) with $z$ as the number of nearest neighbors. The ensemble density in this case can be written as  

\begin{eqnarray}
\rho(H) = \lim_{\sigma \to 0} \; C_a\; \prod_{k=1}^N {\rm e}^{-{H_{kk}^2 \over 2 w^2}}  \;\;  \prod_{k,l=n.n}^N {\rm e}^{-{H_{kl}^2 \over 2 t}}   \; \prod_{k,l \not=n.n}^N {\rm e}^{-{H_{kl}^2 \over 2 \sigma^2}}    
\label{rho5}
\end{eqnarray}

with $C_a$ as the normalization constant.
From eq.(\ref{yw}), the ensemble complexity parameter in this case is \cite{psand}

\begin{eqnarray}
Y \approx  - {1\over  N}  \;  {\rm ln}\left[ |1 -  w^2| \; |1-2 t|^{z/2} \right] + const,
\label{y8}
\end{eqnarray} 

Here the initial state is chosen as a clean lattice with sufficiently far off atoms resulting in zero hopping (i.e both $w=0$ and $t=0$) which corresponds to a localized eigenfunction dynamics with Poisson spectral statistics. (This choice ensures the analogy of initial statistics with the flat band case). Substitution of eq.(\ref{y8}) in eq.(\ref{alm1}) with $\Delta_e(e) = {N \langle I_2 \rangle \over R_1 }$ and $\overline{\langle I_2 \rangle}$ as the typical ensemble as well as spectral averaged IPR at $e$, leads to  

\begin{eqnarray}
\Lambda_{e, AE}(Y,N, e)   = {R_1^2 \over N \; \overline{\langle I_2 \rangle}^2} \; {| \rm ln} (|1-w^2| \; |1-2 t|^{z/2}) |.
\label{alm8}
\end{eqnarray} 

Based on the complexity parameter formulation and verified by the numerical analysis discussed in \cite{psand}, the level density here turns out to be a Gaussian: $R_1(e)={N \over \sqrt{2 \pi \alpha^2}} \; {\rm e}^{-{e^2 \over 2 \alpha^2}}$.
  As indicated by several studies in past  (e.g \cite{mj, emm}), the localization length $\xi$  in this case depends on the dimensionality as well as disorder: (i) $ \xi \approx \pi \; l \approx O(L^0)$  for all $w$  for $d=1$ with $l$ as the mean free path of the electron in the lattice, (ii) $\xi \approx {\rm e}^{{1\over 2} \pi l k_F} \approx O(L^0)$  for all $w$ for $d=2$ with $k_F$ as the Fermi wave-vector and (iii) $ \xi \approx \xi_0(e,w) \; L^{D_2}$  with $D_2={d\over 2}$ for the critical disorder $w=w^*$ for $d>2$. As a consequence,  $\Lambda_e \sim O(1/N)$ for $d \le 2$ which implies the statistics approaching an insulator limit a $N \to \infty$.  For $d >2$, $\Lambda_e$  in the spectral bulk is size-independent only for $w=w*$ (for a  fixed $t$), thus indicating only one critical point \cite{psand} of transition from delocalized to localized states with increasing disorder.  

An important point worth re-emphasizing is here is that notwithstanding the $N$-dependence of $Y-Y_0$  same  for AE and the flat bands (discussed in section II.a), the statistics of energy levels and eigenfunctions in the two cases undergoes an inverse  transition. This occurs because  
$\Lambda_e$, the only parameter governing the spectral statistics, depends on the localization length and mean level density which have different response to disorder $w$ in the two cases.


\vspace{0.1in}

{\bf Rosenzweig Porter (RP) Ensemble:}  This represents an ensemble of Hermitian matrices with independent, Gaussian distributed  matrix elements with zero mean, and different variance for the diagonals and the off diagonals. The ensemble density $\rho(H)$  in this case can be given as

\begin{eqnarray}
\rho(H)  \propto 
{\rm exp}{\left[-\frac{1}{2} \; \sum_{i=1}^{N} H_{ii}^2 -  
 (1+\mu_0) \sum_{i,j=1; i < j}^{N} H_{ij}^2 \right]}  
\end{eqnarray}
As clear from the above, contrary to multi-parametric dependent Anderson case, the RP ensemble depends on the single parameter  i.e ratio of the diagonal to off-diagonal variance (besides matrix size).

The ensemble density given above is analogous to the Brownian ensemble (BE) which arises due to a single parametric perturbation of an ensemble of diagonal matrices by a GOE ensemble (discussed in detail in section 2 of \cite{psand}  and also in \cite{psbe}). Clearly the statistics of BE or RP ensemble lies between Poisson and GOE limits and  depends  on a single parameter which can be given as follows. The choice of initial condition as an ensemble of diagonal matrices (which corresponds to $\mu_0 \rightarrow \infty$) gives $Y-Y_0 = {1\over 4 \mu_0}$ (see eq.(11)  of \cite{psand}, also can be seen from eq.(\ref{y}) by substituting $v_{kl;q}=\delta_{kk}+{(1-\delta_{kk})\over 2(1+\mu_0)}\delta_{q1}$, $b_{kl,q}=0$ for all $k,l$-pairs) which leads to 

\begin{eqnarray}
\Lambda_{e,BE}(e)=\frac{Y-Y_0}{\Delta_e(e)^2} = \frac{R_1^2}{ 4 \mu_0 }.
\label{alm9}
\end{eqnarray}
Note, the 2nd equality in the above equation is obtained by using $\Delta_e(e)= {1\over R_1(e)}$ (see \cite{loc} for a brief explanation).

As discussed in \cite{psbe}, the size-dependence of $R_1(e; \mu_0)$ for a BE or RP ensemble changes from $\sqrt{N}$ to $N$. This in turn indicates the existence of two critical points: (i) for 
{\bf $\mu_0=c_1 N^2$}:  here $R_1= {N\over \sqrt{ \pi}}{\rm e}^{-e^2}$ which gives  $\Lambda_{e, BE}={1\over 4 \pi c_1}{\rm e}^{-2 e^2}$,    
(ii)  for {\bf $\mu_0=c_2 N$}: here  $R_1(e)= (b \pi)^{-1} \; \sqrt{2 b N- e^2}$ leading to 
$\Lambda(e)=\frac{2 b N-e^2}{\pi^2 b^2 N c_2}$ with $b \sim 2$.  The two critical points here corresponds to a transition from localized $\to$ extended  $\to$ delocalized states with decreasing $\mu_0$ \cite{psbe}.  


\vspace{0.1in}

{\bf Parametric values for the analogues:}
For numerical analysis of Anderson ensemble,  we consider a three dimensional cubic lattice  with hard wall boundary conditions, on-site Gaussian disorder $w$ and a random hopping with $t=1/12$. For Brownian ensemble, we choose the case with $\mu_0=c \; N^2$; (note the latter choice is arbitrary). The system parameters  for the Anderson and Brownian ensemble analogs of a weakly disordered flat band can now be obtained by  invoking following condition 

\begin{eqnarray}
\Lambda_{e, FE}=\Lambda_{e, AE}=\Lambda_{e, BE}. 
\label{almu}
\end{eqnarray}
with $\Lambda_{e, FE}$, $\Lambda_{e, AE}, \Lambda_{e, BE}$ given by eq.(\ref{alm10}), eq.(\ref{alm8}) and eq.(\ref{alm9})  respectively. 


Figure 6 displays a comparison of the nearest neighbor spacing distribution for two cases of disordered chequered board lattice (with Fermi energy in bulk of the flat band) with AE and BE analogues predicted by eq.(\ref{almu}). The numerically obtained values for the analogs near ($e \sim 0$ for each case) are as follows:

(i) {\it weak disorder analogy:} (a) FE: $N=1156$, $w^2=0.01$, $\varepsilon=2$, $t=1$, $\langle I_2^{typ}\rangle=0.0116$, $R_1(e)={0.248 \; N \over w}$ which gives $\Lambda_{e, FE}=0.395$, (b) AE: $N=512$, $w^2=4.15/6$, $t=1/12$, $z=6$, $\langle I_2^{typ}\rangle=0.025$, $R_1=0.2983 \times N$ with $\Lambda_{e, AE}=0.465$ and
(c) BE: $N=512$, $c=0.2$ with $\Lambda_{e, BE}=0.398$.

(ii) {\it strong disorder analogy:} 
(a) FE: $N=1156$, $w^2=10$, $\varepsilon=2$, $t=1$, $\langle I_2^{typ} \rangle=0.1149$, $R_1(e)={0.3 \; N\over w}$ which gives $\Lambda_{e, FE}=1.3 \times 10^{-3}$, (b) AE: $N=512$, $w^2=120.15/6$, $\langle I_2^{typ} \rangle \approx 0.3$, $R_1 \approx 0.1/N$ with $\Lambda_{e, AE}=7.58 \times 10^{-4}$
and (c) BE: $N=512$, $c=69.2$ with $\Lambda_{e, BE}=1.15 \times 10^{-3}$.

The AE and BE analogs for the other flat band cases can similarly be obtained.  Alternatively, statistics of the perturbed flat band considered here can also be mapped to the AEs with different system conditions and the BE with $\mu_0 \propto N$. To confirm that this analogy is not a mere coincidence and exist for other $\Lambda_e$ values too, we compare  these ensembles for full crossover from $\Lambda_e =0 \to \infty$.  One traditionally used measure in this context is the relative behavior of the tail of nearest-neighbor 
spacing distribution $P(s)$, defined as 
\begin{eqnarray}
\gamma (\delta; \Lambda) = \frac{\int_0^\delta (P(s; \Lambda)-P(s; \infty))
{\rm d}s} {\int_0^\delta (P(s; 0)-P(s; \infty)){\rm d}s}
\label{alp}
\end{eqnarray}
with $\delta$ as one of the two crossing points of $P_o(s) = P(s; \infty)$ and $P_p(s) = P(s; 0)$ 
(here the subscripts $o$ and $p$ refer to the GOE and Poisson 
cases respectively) \cite{mj, psand}.  As obvious, $\gamma=0$ and $1$ for GOE  and Poisson limit respectively  and a fractional value of $\gamma$ indicates the probability of small-spacings  different from the two limits. In limit $N \to \infty$, a $\gamma$ value different from the two end points is an indicator of a new universality class of statistics and therefore a critical point.    Figures 6(c) and 6(d)  show a comparison of $\gamma$ for two $\delta$-values for three systems: $\gamma_1 =\gamma(\delta_1)$ and $\gamma_2=\gamma(\delta_2)$, with $\delta_1=0.4699, \delta_2=1.9699$; the display confirms our theoretical claim regarding the analogy of the three systems. 
It must be noted that the $\Lambda_e $ for FE never approaches a value as large as that of AE and BE;  following from eq.(\ref{alm3}) and eq.(\ref{alm5}), it first increases and then decreases beyond a disorder-strength $w \sim 1$. This is contrary to AE and BE for  which $\Lambda_e$ decreases with increasing disorder. This behavior is also confirmed by our numerical analysis displayed in figure.

\section{conclusion}


In the end we summarize with  main insights and results given by our  analysis. We find that a disordered system, with one or more flat bands in clean limit, can undergo two types of localization to delocalization transition. 
In weak disorder regime  (below a system specific disorder strength, say $w_c$), the localization  is insensitive to disorder strength and persists even for a very small disorder. This in turn leads to a critical spectral statistics, disorder-independent  and  analogous to a Brownian ensemble  intermediate to Poisson and Wigner-Dyson classes. But in strong disorder regime ($w > w_c$), the behavior is analogous to that of a disorder-driven, standard Anderson transition (for single particle bands) or many body localization transition (for many particle  bands) in which  a size-invariant spectral statistics occurs only at  specific  disorder strengths; the statistics here  is again  analogous to a Brownian ensemble but characterized by a different parameter value. The clearly reveals the influence the underlying  scattering has on the transitions in the two regimes: although it affects the  transition parameter dependence on disorder, the spectral statistics in both regimes belongs to one parameter dependent universality class  of Brownian ensembles.

The analysis presented here is based on a single parameter formulation of the  spectral statistics. This not only helps in theoretical understanding of the numerical results given  by our as well as previous studies \cite{nmg2} but also reveals new features. For example, it provides a unified formulation of the spectral statistics in the weak and strong disorder regimes (notwithstanding different scattering conditions). It also identifies the spectral complexity parameter as the transition parameter  and leads to  its exact mathematical expression  which in turn helps in the search of criticality  in a disorder perturbed flat band; this occurs when the system conditions conspire collectively to render the spectral complexity parameter size-independent. More clearly, the criticality requires the ensemble complexity parameter, an indicator of the average uncertainty in the system, measured in the units of local mean level spacing, to become scale-free.  The underlying localization dynamics clearly leaves its fingerprints on the  transition parameter; the latter  turns out to be  disorder independent in weak disorder regime but is disorder dependent in strong disorder regime.

The advantage of complexity parameter based analysis goes beyond a search for criticality in perturbed flat bands. It also  reveals an important analogy in the localization to delocalization crossover in finite systems:  notwithstanding the difference in the number of critical points as well as equilibrium limits, the statistics of a disordered flat band can  be mapped to that of a single parametric Brownian ensemble \cite{psbe} as well as multi-parametric Anderson ensemble \cite{psand} (see section VI). The analogy of these ensembles to other multi-parametric ensembles intermediate between Poisson and Wigner-Dyson is already known \cite{ssps, ps-all, psvalla, psbe}. In fact it seems a wide range of localization $\to$ delocalization transition can be modeled by a single parameter Brownian ensemble appearing between Poisson and GOE (Rosenzweig-Porter ensemble) \cite{psrp}. This hints at a large scale universality and a hidden web of connection underlying complex systems even for partially localized regime. 
Note the universality of spectral statistics and eigenfunctions  in  ergodic or delocalized waves regime is already known but the complexity parameter formulation reveals a universality  even at the critical point of widely different systems (of same global constraint class) if their complexity Parameters are equal. It is relevant for the following reason: it is well known that average properties of systems often show a power law behavior at the critical point and can be classified into various universality classes based on  their powers, referred as the critical exponents. However, in case of a complex system where the fluctuations of physical properties are often comparable to their averages, it is not enough to know the universality classes of critical exponents. An important question in this context is whether there are universality classes among the fluctuation properties too?  As discussed in section VI, such universality classes can indeed be identified based on the complexity parameter formulation. This issue will be discussed in more detail in a future publication.

Our study gives rise to many new queries. For example, an important question is whether weak particle-particle interactions in clean flat bands can  mimic the role of weak disorder in the perturbed flat bands. At least the complexity parameter formulation predicts this to be the case but a thorough investigation of the fluctuations is needed to confirm the prediction.   A detailed analysis of the role of the symmetries in flat band physics using complexity parameter approach still remains to be investigated.  Our analysis seems to suggest the existence of a mobility edge too however this requires a more thorough investigation. We expect to explore some of these questions in future.

\vfill\eject

\oddsidemargin=-10pt
\begin{figure}
\centering
\includegraphics[width=\textwidth, height=\textwidth]{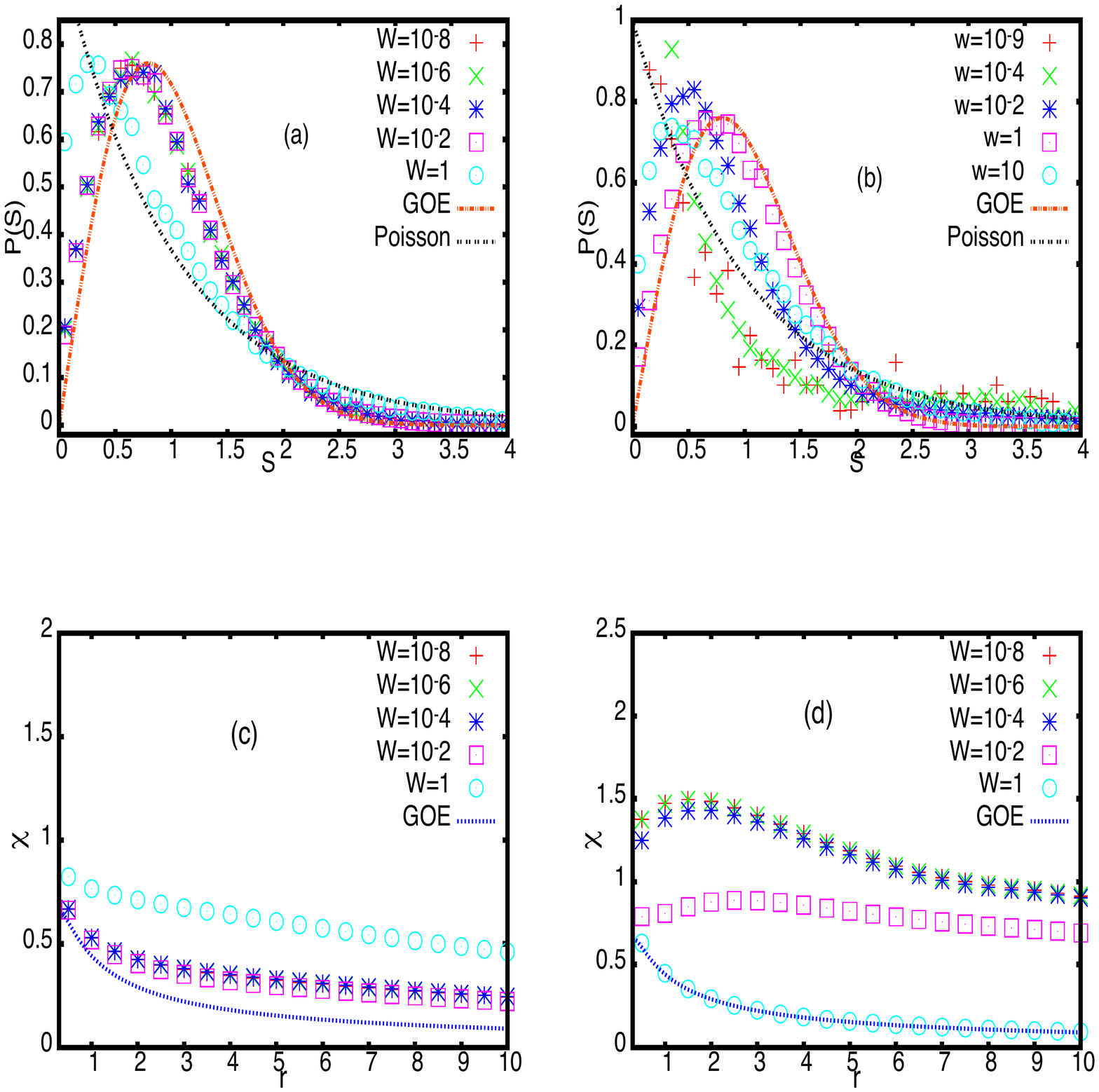} 
\vspace*{-30 mm}
\caption{ 
{\bf Disorder dependence of  spectral measures in two energy ranges:} 
(a) $P(S)$ in the bulk of flat band  ($e \sim 0$),
 (b) $P(S)$ in the bulk of dispersive band ($e \sim 4$), 
(c)  $\chi(r)$  in the bulk of flat band ($e \sim 0$),
(d)  $\chi(r)$ in the bulk of dispersive band ($e \sim 4$).
Here  $W=w^2$ and $P(S)$ refers to the distribution of the nearest-neighbor spacing $S$ and $\chi(r)$ as the spectral compressibility for the unfolded eigenvalues taken from a narrow energy-range around the specific energy  for a fixed system size $L=70$. The total number of eigenvalues used in each case is approximately $10^5$.  As clear from parts (a, c), the statistics is near GOE and disorder-insensitive for $w <1$ but approaches Poisson limit for $w>1$. Clearly, with $w=0$ as the Poisson case (due to degeneracy in flat band spectrum), increasing disorder from zero leads to a change of statistics from Poisson to near-GOE to Poisson, which corresponds 
to a localization-delocalization-localization crossover of the eigenfunctions in the bulk of the flat band. But parts (b,d) indicate a disorder insensitivity as well as inverse crossover in  the dispersive band: with $w=0$ as GOE case, increasing disorder from zero leads to a change of statistics from GOE $\to$ Poisson $\to$ GOE which corresponds to a delocalization-localization-crossover of the eigenfunctions in the bulk of the dispersive band. 
}
\label{fig1}
\end{figure}

\begin{figure}
\includegraphics[width=1.\textwidth, height=\textwidth]{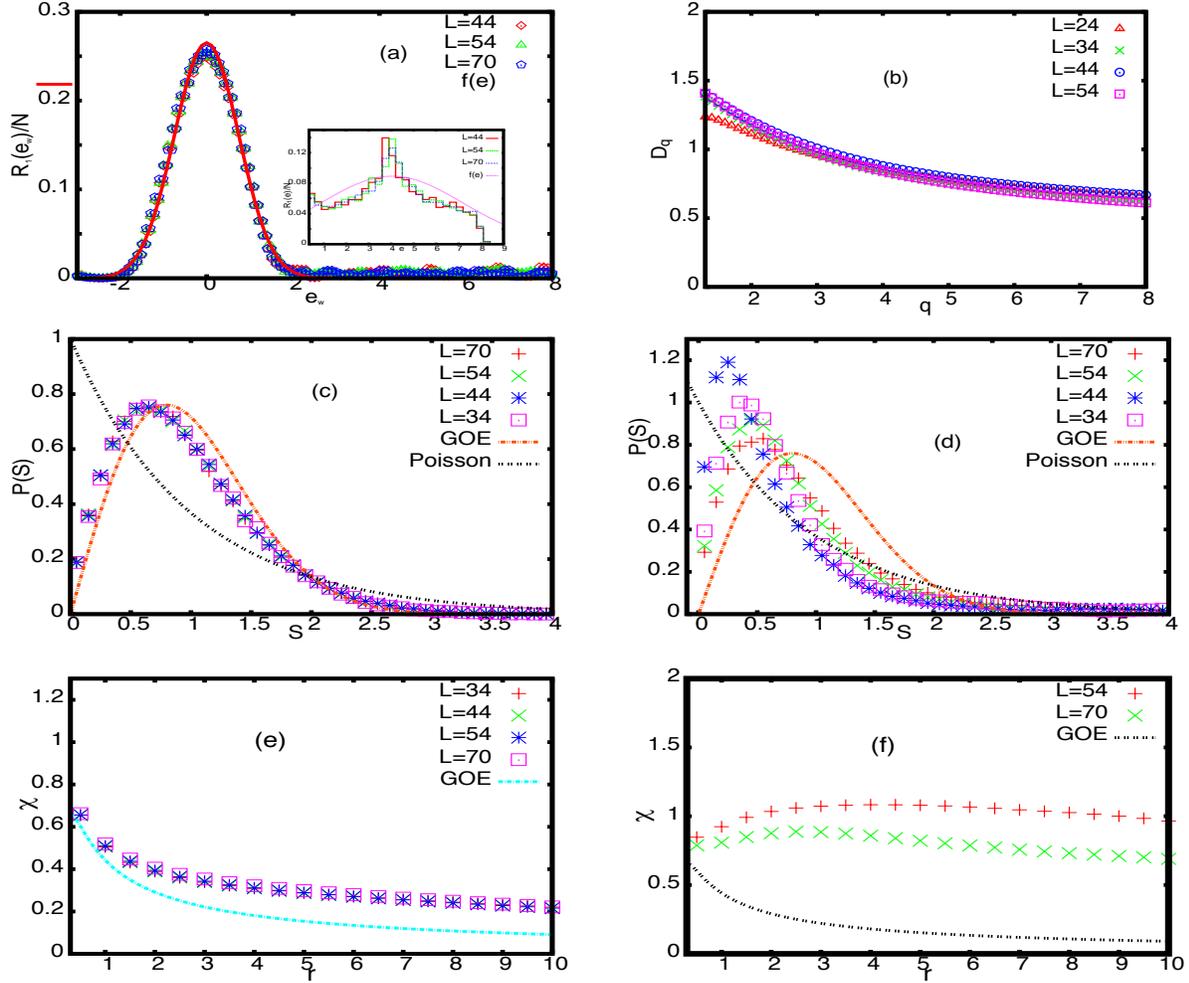} 
\vspace*{-30 mm}
\caption{ 
{\bf Critical spectral statistics for weak  disorder $w=\sqrt{3} \times 10^{-5}$:}
(a) Level density  in the flat band (inset showing the behavior in the dispersive band) with fit $f(e_w)={1 \over 2 \sqrt{1.25 \pi} }\; {\rm e_w}^{-0.8 \; e_w^2} $, 
(b) $D_q$ in the flat band ($e \sim 0$),
(c) $P(S)$ for flat band bulk ($e \sim 0$), (d) $P(S)$ for dispersive band bulk ($e \sim 4$), 
(e) $\chi(r)$ for flat band bulk ($e \sim 0$) ,  
(f) $\chi(r)$ for dispersive band bulk ($e \sim 4$), 
The parts (c-e) also display the GOE and Poisson limits. 
Here, with $\langle {\mathcal I}_2 \rangle=0.0116$ and $R_1 \approx {0.248 \; N \over w}$, eq.(\ref{alm10}) gives $\Lambda_e=0.395$ near $e \sim 0$. 
The convergence of the curves for different sizes in parts (c, e) indicates
scale-invariance of the statistics in the flat band. The behavior is critical due to $P(S)$ being different from the two end-points, namely, Poisson and GUE statistics even in large size limit. This is also confirmed by the $\chi$-behavior shown in part (e), approaching a constant value $0..2$ for large $r$, and  $D_q$ behavior shown in part (b). Note the $\chi$-value is in agreement with eq.(\ref{ch2}) and $D_2$ is consistent with relation $D_2=d(1-2 \chi)$ with $d=2$.   
The survival of scale-invariance and partially localized behavior 
even for such a weak disorder indicates the critical point of
the inverse Anderson transition to occur at zero disorder strength. 
In contrast, parts (d) and (f) indicate that the bulk statistics in the dispersive band is size-dependent and is not critical. 
}
\label{fig2}
\end{figure}

\oddsidemargin=-10pt
\begin{figure}
\centering
\includegraphics[width=1.\textwidth, height=\textwidth]{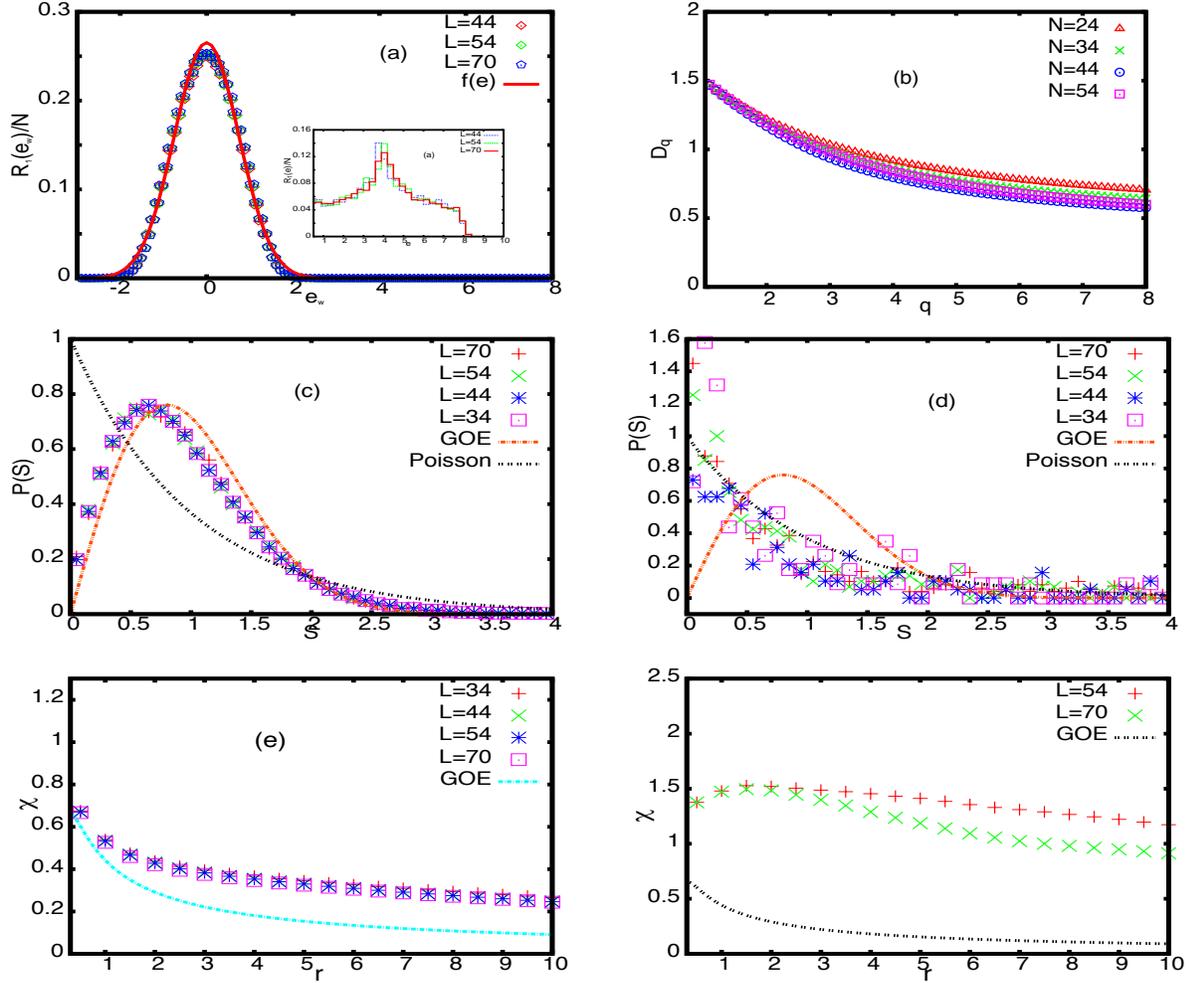} 
\vspace*{-30 mm}
\caption{
{\bf Critical spectral statistics for weak  disorder:  $w=0.1$:}
The details here are same as in figure 2; the fit in part (a) is $f(e_w)={1 \over 2 \sqrt{1.25 \pi}} \; {\rm e}^{-0.8 \; e_w^2} $. Here, with $\langle {\mathcal I}_2 \rangle=0.0116$ and $R_1(e) \approx {0.248 \; N\over w}$, eq.(\ref{alm10}) gives $\Lambda_e \approx 0.395$. 
 As can be seen from parts (b) and (e)  $\chi=0.2, D_2=1.2$ near $e \sim 0$ which is again  in agreement with eq.(\ref{ch2}) as well as relation $D_2=d(1-2 \chi)$ with $d=2$.   
The analogy of the statistics with the case displayed in figure 2 indicates the disorder insensitivity of the statistics for $w <1$.
 }
\label{fig3}
\end{figure}

\oddsidemargin=-10pt
\begin{figure}
\centering
\includegraphics[width=1.\textwidth, height=\textwidth]{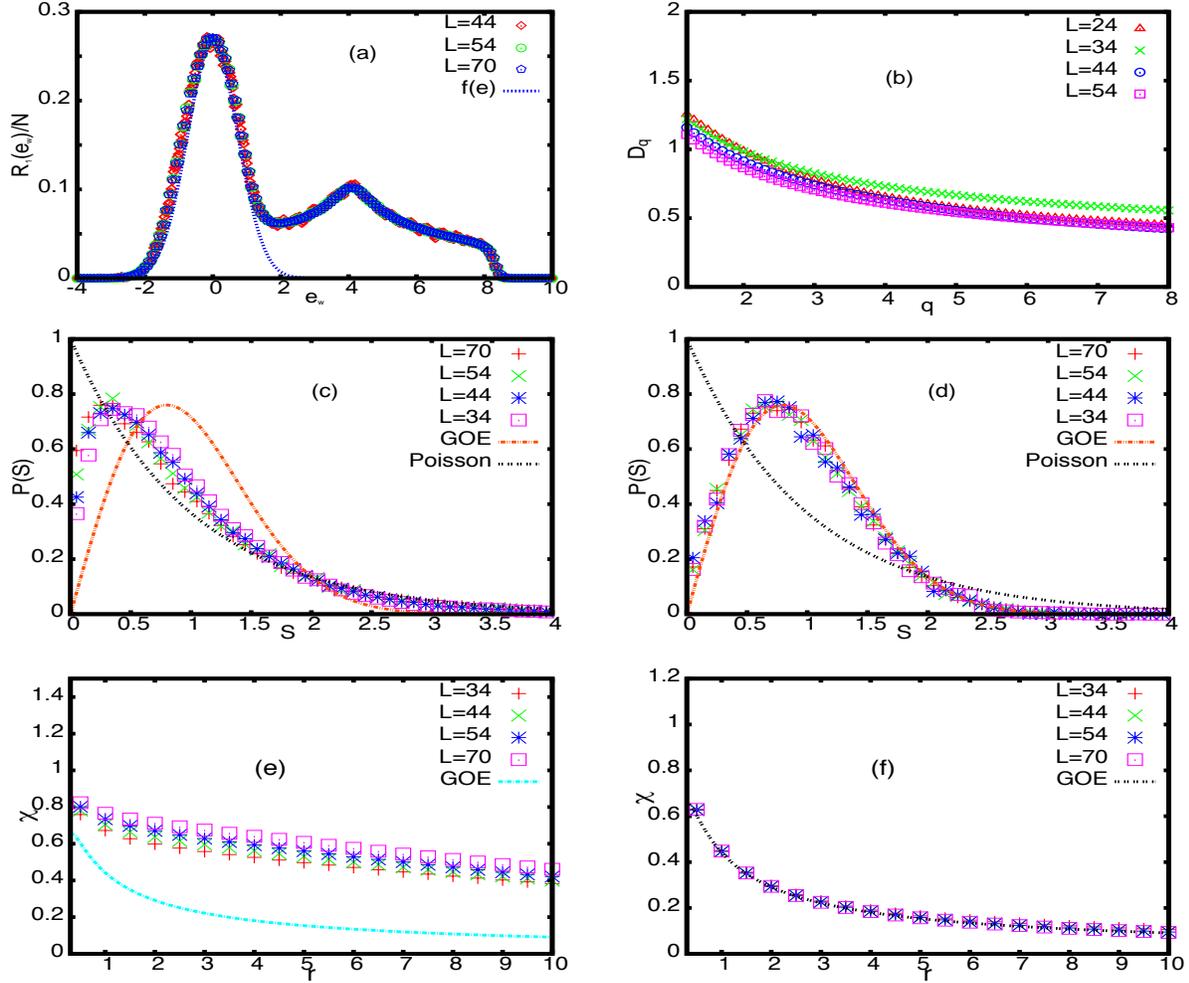} 
\vspace*{-30 mm}
\caption{{\bf Critical spectral statistics for  disorder:  $w=1$:}
(a) Level density $R_1(e_w)/N$ along with fit $f(e_w)={1 \over 2 \sqrt{1.15 \pi} }\; {\rm e}^{-0.76 \; e_w^2} $ with $e_w=e/w$, 
(b) $D_q$ near $e \sim 0$,
(c) $P(S)$ near $e \sim 0$, (d) $P(S)$ near $e \sim 4$, 
(e) $\chi(r)$ near $e \sim 0$,  
(f) $\chi(r)$ near $e \sim 4$. 
Here, with $\langle {\mathcal I}_2 \rangle=0.0269$ and $R_1(e) \approx {0.261 \; N\over {w}}$, eq.(\ref{alm10}) gives $\Lambda_e=0.08$ near $e \sim 0$. 
The parts (b) and (e)  now give $D_2=0.95, \chi =0.5$ near $e \sim 0$;  this values are no longer consistent with eq.(\ref{ch1}) or eq.(\ref{ch2}) or relation $D_2=d(1-2 \chi)$; the latter is however expected because the $D_2-\chi$ relation is expected to be valid only for small $\chi$. 
Further, as can be seen from part (a), the two bands start merging at this disorder strength. 
In contrast to weak disorder case, the statistics in flat and dispersive bands are now reversed i.e 
closer to Poisson and GOE respectively.
But the survival of scale-invariance and partially localized behavior 
even for this disorder strength indicates the dominance of flat band spectrum on that of the dispersive band.
}
\label{fig4}
\end{figure}

\oddsidemargin=-10pt
\begin{figure}
\centering
\includegraphics[width=1.\textwidth, height=\textwidth]{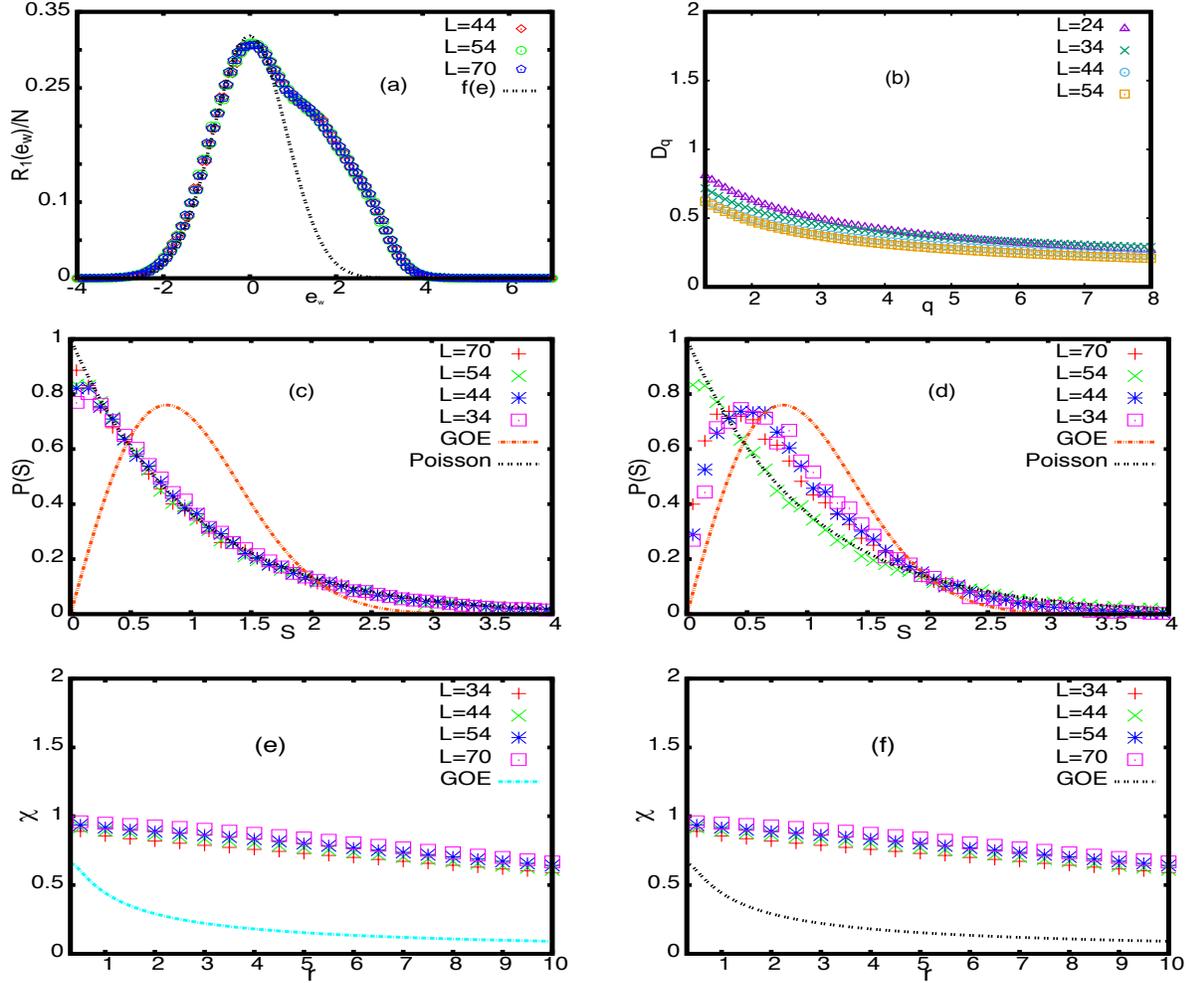} 
\vspace*{-30 mm}
\caption{{\bf Critical spectral statistics for strong  disorder $w=\sqrt{10}$:}
(a) Level density $R_1(e_w)/N$ along with fit $f(e_w)={1 \over 2 \sqrt{0.82 \pi}} \; {\rm e}^{-1.2 \; e_w^2} $ with $e_w=e/w$,
(b) $D_q$ near $e \sim 0$,
(c) $P(S)$ near $e \sim 0$, (d) $P(S)$ near $e \sim 4$, 
(e) $\chi(r)$ near $e \sim 0$,  
(f) $\chi(r)$ near $e \sim 4$. 
Here with $\langle {\mathcal I}_2 \rangle=0.1149$ and $R_1(e) \approx 0.3 \; N/{w}$, eq.(\ref{alm10}) gives $\Lambda_e=1.38 \times 10^{-3}$ near $e \sim 0$.
As can be seen from parts (b) and (e)  $\chi \approx 0.6, D_2=1.2$ near $e \sim 0$. Again there is no agreement with eq.(\ref{ch1}) or relation $D_2=d(1-2 \chi)$. Clearly eq.(\ref{ch1}) seems to be applicable for a much smaller $\Lambda_e$. Further a large $D_2$ value here seems to be the effect of the complete merging between two bands  giving rise to a 
new band. The statistics approaches Poisson regime and an intermediate regime for $e \sim 0$ and $e \sim 4$ respectively, 
both indicating  localized dynamics of wave functions. The system has now reached an insulator limit in the bulk energies but is still 
partially localized at the edge of the new band.}
\label{fig5}
\end{figure}

\oddsidemargin=-10pt
\begin{figure}
\centering
\includegraphics[width=1.\textwidth, height=1.2\textwidth]{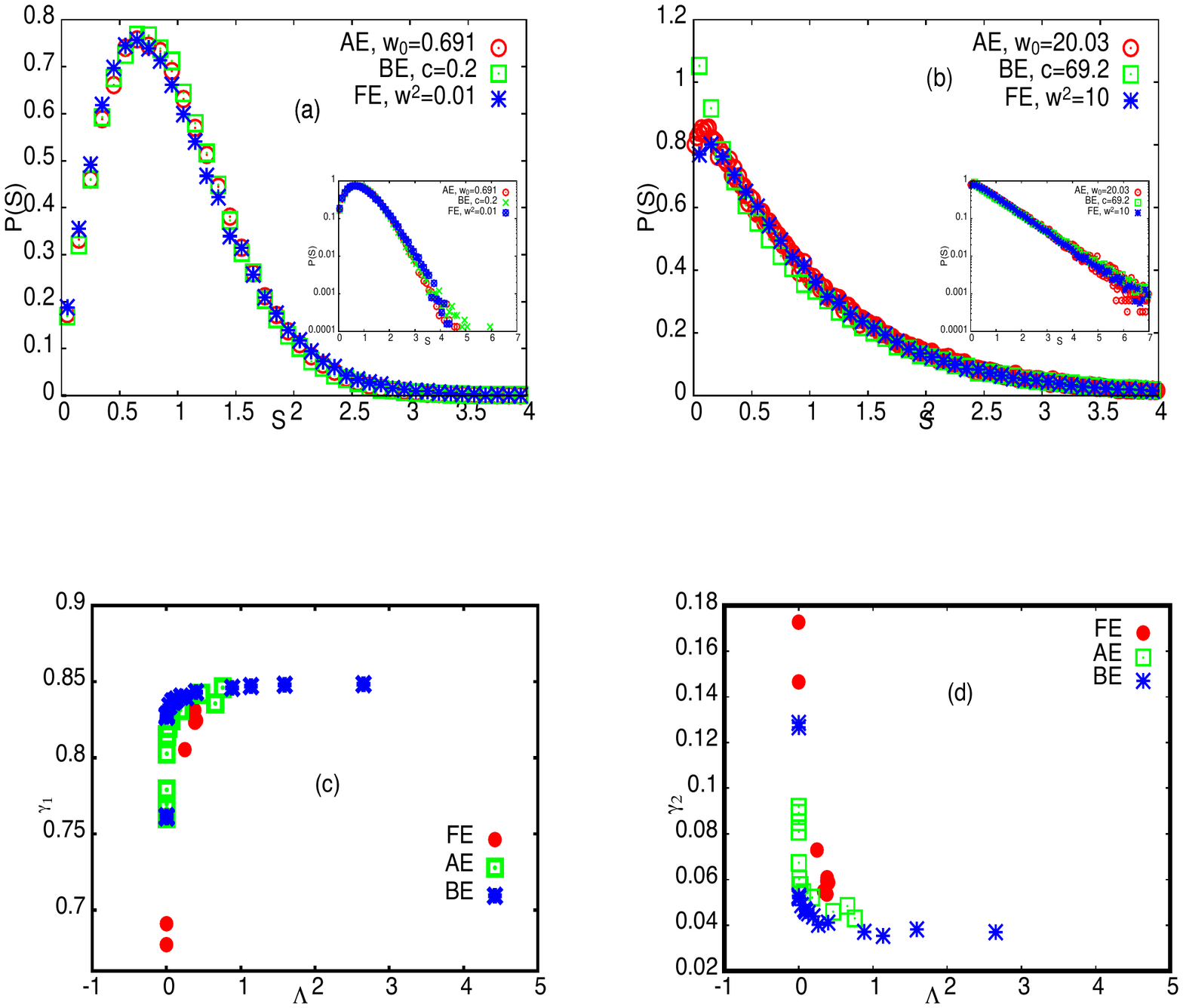} 
\vspace*{-60 mm}
\caption{ 
{\bf Comparison of Flat band ensemble (FE) with AE and BE:}  
Here the parts (a) and (b) display the $P(s)$ comparison for the AE, BE analogues of a weakly disordered flat band for two disorders $w^2=0.1$ and $10$. The AE and BE analogues have been obtained by the conditions $\Lambda_{e, FE}=\Lambda_{e, AE} = \Lambda_{e, BE}$ given by eq.(\ref{alm10}), eq.(\ref{alm8}) and eq.(\ref{alm9})  respectively; the system parameter for the three ensembles leading to approximately same $\Lambda_e$ near $e \sim 0$ are as follows: 
(a) {\bf FE:} $N=1156, w^2=10^{-2}, \varepsilon =2, t=1$, {\bf AE:} $N=512, w^2={4.15\over 6}, t={1\over 12}$, {\bf BE:} $N=512, c=0.2$, and, 
(b) {\bf FE:} $N=1156, w^2=10, \varepsilon=2, t=1$, {\bf AE:} $N=512, w^2={120.15 \over 6}, t={1\over 12}$, {\bf BE:} $N=512, c=69.2$.
Tor rule out the accidental coincidence, we also compare $\gamma_1=\gamma(0.4699), \gamma_2=\gamma(1.9699)$  for a range of $\Lambda_e$ values. The results are displayed in parts (c) and (d), respectively. As clearly visible from the figures, the values for all three ensemble collapse on the same curve for small $\Lambda_e$ values. But while $\Lambda_e$ of a flat band decreases for both small and large $\Lambda_e$, the $\Lambda_e$ of an AE and BE smoothly increases from $0$ to a large value with decreasing disorder. The  deviation in their behavior for large $\Lambda_e$ values is therefore an indicator of the different nature of transition.    
}
\label{fig6}
\end{figure}


\begin{thebibliography}{10}
\bibitem{pp2}
The matrix representation of an operator depend on the choice of the basis 
and the constraints on the dynamics. This in turn also affects the behavior of 
the eigenfunctions (due to their basis-dependence).    
Previous studies claiming multifractal eigenfunctions at a critical point are often 
concerned with Hamiltonian matrices with Hermiticity  as the only constraint. For cases where 
the system is subjected to additional constraints e.g.those resuling in positive semidefinite
matrices as in \cite{gu1}, further investigations are needed to confirm the mutifractality.
\bibitem{oh1}
 The formulation presented here is applicable for an arbitray  basis as long as the matrix
remains Hermitian.  
However the choice of a basis which preseves global constraints of the system i.e the symmetry and 
conservation laws can provide physically relevant information; for example, 
it permits a comparison of different complex systems (those with same global constraints) and reveals 
their statistical analogies through complexity parameter formulation.
\bibitem{pp3}
The term “partially localized state” here refers to an eigenstate of the system which extends upto a few basis states only, say $n$, with $1 < n < N$  in a  $N$-dimensional basis. It can then be characterized by an inverse participation ratio $I_2$ between zero and $1/N$. A fully localized state with this definition corresponds to the one confined to a single basis state and has $I_2 =1$. Similarly a delocalized state corresponds to the one 
                                           extended over all basis states and has $I_2=1/N$.
\bibitem{loc}
The  relation $\Delta_e(e)={1\over  \rho(e) \; \xi^d}$ is applicable only for the sparse matrices and not in case of the dense matrices. This can be explained as follows. Consider the matrix representation of an operator $H$ in an arbitrary  basis $|k \rangle$, $k=1 \to N$ in which all off-diagonals are of the same order (as in case of a BE). The  matrix element, say, $H_{kl}$ can  be written as $H_{kl} = \sum_{n=1}^N e_n \; U_{kn} \; U^*_{ln}$ with $U_{kn}$ as the $k^{th}$ component of the eigenfunction $U_n$ corresponding to the eigenvalue $e_n$ of $H$. For cases where $H$ is a random matrix, $e_n$ are random variables too. Clearly, for $H_{kl}$ to be of the same order for all basis pairs $k,l$ for arbitrary $e_1,\ldots, e_N$, the correlation $U_{kn} U^*_{ln}$ should typically be of the same order (i.e almost independent of $k,l$ although it may depend on $n$). This requires a state $U_n$ to typically have almost same overlap with all basis states, thus implying an extended state in the volume $L^d$ which in turn leads to $\Delta_e(e)={1\over  \rho(e) \; L^d }={1\over R_1}$.
\end{thebibliography}
 \end{document}